\documentclass[english,aps,nofootinbib,prd,preprint,showpacs,fleqn]{revtex4}
\usepackage[T1]{fontenc}
\usepackage[latin1]{inputenc}
\usepackage{color}
\usepackage{graphicx}
\usepackage{amssymb}

\makeatletter

\providecommand{\LyX}{L\kern-.1667em\lower.25em\hbox{Y}\kern-.125emX\@}

\usepackage{mathrsfs} 
\renewcommand{\vec}[1]{{\bf #1}}

\usepackage{babel}
\makeatother
\begin{document}
\def\STAR{{\sc Star }}

\def\PHENIX{{\sc Phenix }}

\newcommand{\Lagrangian}{{\mathcal{L}}_{int}}

\newcommand{\reaction}{A+B\, \rightarrow \, (V^{*}\rightarrow \, l_{1}+\bar{l}_{2})+X}

\newcommand{\wh}[1]{\widehat{#1}}

\newcommand{\wt}[1]{\widetilde{#1}}

\newcommand{\shat}{\widehat{s}}

\newcommand{\that}{\widehat{t}}

\newcommand{\uhat}{\widehat{u}}

\newcommand{\GHAHB}{G_{j\bar{k}}^{(H_{A},H_{B})}(\theta ,\varphi )}

\newcommand{\sigt}{\tilde{\sigma }}

\newcommand{\Tqq}{T_{q\bar{q}}}

\newcommand{\TqG}{T_{qG}}

\newcommand{\TGq}{T_{Gq}}

\newcommand{\MSbar}{\overline{MS}}

\newcommand{\A}{{\mathcal{A}}}

\newcommand{\eps}{\varepsilon }

\newcommand{\B}{{\mathcal{B}}}

\newcommand{\C}{{\mathcal{C}}}

\renewcommand{\floatpagefraction}{1} \renewcommand{\S}{{\cal S}}\renewcommand\vec[1]{{\bf #1}}

\preprint{hep-ph/0304001v.2}


\title{Soft parton radiation in polarized vector boson production:\\
theoretical issues}

\author{P.M. Nadolsky\protect$^{1}$}

\email{nadolsky@mail.physics.smu.edu}

\author{C.-P. Yuan\protect$^{2}$}

\email{yuan@pa.msu.edu}

\affiliation{$^{1}$Department of Physics, Southern Methodist University, \\
Dallas, Texas 75275-0175, U.S.A.\\
$^{2}$Department of Physics \& Astronomy, Michigan State University,
East Lansing, Michigan 48824, U.S.A.}

\date{June 24, 2003}

\begin{abstract}
Accurate measurement of spin-dependent parton distributions in production
of electroweak bosons with polarized proton beams at the Relativistic
Heavy Ion Collider depends on good understanding of QCD radiation
at small transverse momenta $q_{T}$ of vector bosons. We present
a theoretical formalism for small-$q_{T}$ resummation of the cross
sections for production of $\gamma ^{*}$, $W^{\pm }$, and $Z^{0}$
bosons, with the subsequent decay of these bosons into lepton pairs,
for arbitrary longitudinal polarizations of the proton beams. 
\end{abstract}
\maketitle
\newpage

\section{\label{sec:Intro}Introduction}

The commissioning of polarized proton beams with energies from 100
to 250 GeV at the Relativistic Heavy Ion Collider (RHIC) \cite{RHIC2001}
has started a new stage in physics of spin-dependent collisions. In
particular, measurement of spin-dependent parton distribution functions
(PDFs) will be the top priority in the experiments with longitudinal
polarization. So far, these spin-dependent counterparts of the unpolarized
PDFs were constrained only by the lepton-nucleon deep inelastic scattering
(DIS) data \cite{EMC1,EMC2,SMC,E142,E143,E154,E155,HERMES:polDIS96,HERMES:polDIS98},
which concentrates at relatively small momentum transfers ($Q^{2}\sim 1\mbox {\, GeV}^{2}$).
In contrast, RHIC will explore the spin-dependent PDFs in a much larger
range of $Q^{2}$ and using a variety of spin-dependent particle reactions,
including production of Drell-Yan lepton pairs and resonant production
of $W^{\pm }$ and $Z^{0}$ bosons \cite{RHICOverview}. 

Vector boson production (VBP) with polarized hadron beams, which proceeds
through annihilation of a quark and antiquark at the Born level, is
the most natural candidate to probe spin-dependent quark polarizations
\cite{Close:1977qx,Bourrely:1980mr,Craigie:1983tk}. Furthermore,
in $pp$ collisions the Born level cross sections are sensitive to
the distributions of sea quarks. In this sense, VBP complements production
of jets, pions, and heavy quark flavors, which primarily probe the
gluon distribution. No wonder that extensive bibliography is dedicated
to spin-dependent production of Drell-Yan pairs \cite{Ratcliffe:1983yj,Richter-Was:1985jp,Cheng:1990hf,Bourrely:1991pz,Weber1,MathewsRavindran,Chiappetta:1993cp,Kamal1,Gehrmann1,Kumano:1999bt,Dressler:1999zv,Gluck:2000ek,Ravindran:2002na,Kodaira:2003tq}
and massive electroweak bosons \cite{Chiappetta:1985dy,Weber2,Bourrely:1994sc,Bourrely:1993dd,Nadolsky:1995nf,Bourrely:1995fw,Kamal2,Gehrmann2,Gluck:2000ek}.

The production of $W^{\pm }$ bosons presents a particularly interesting
opportunity to learn about the quark spin structure due to the maximal
violation of space-reflection parity in the $q\bar{q}W$ coupling
and non-trivial mixing of the quark flavors through the Cabibbo-Kobayashi-Maskawa
(CKM) matrix \cite{Cabibbo:1963yz,Kobayashi:1973fv}. While the former
feature allows non-vanishing single-spin cross section asymmetries
(which are simpler than the parity-conserving double-spin asymmetries),
the latter feature facilitates the study of the flavor dependence
of the sea quark PDFs. The issue of flavor symmetry breaking in the
polarized quark sea was brought in the limelight recently by the results
on semi-inclusive hadroproduction in spin-dependent DIS by HERMES
Collaboration \cite{HERMES:polSIDIS99}. In that measurement, the
issues of higher-order and higher-twist corrections are an important
consideration due to small values of $Q^{2}$. At the same time, $W$
boson production at RHIC has a potential to pin down the quark sea
flavor structure at large $Q^{2}$, i.e., at energies where perturbative
quantum chromodynamics (PQCD) is truly valid. 

The original method \cite{Bland:1999gb,RHICOverview} for the measurement
of polarized PDFs in $W$ boson production at RHIC relies on the reconstruction
of the spin-dependent distribution $d\sigma /dy$ over the rapidity
$y$ of the $W$ bosons. The method is based on the observation that
the Born level single-spin asymmetry of $d\sigma /dy$ is described
by a simple theoretical expression, which further reduces to the ratio
$\Delta q(x)/q(x)$ of the polarized and unpolarized PDFs when the
absolute value of $y$ is large. Unfortunately, the application of
this method is obstructed by specifics of the detection of the $W^{\pm }$
bosons at RHIC. Since neither of RHIC detectors controls the energy
balance in particle reactions,  the energy and momentum of the neutrino
in the decay are completely unknown, so that the information about
the momentum of the $W$ boson is incomplete. As a result, the determination
of the rapidity $y$ of the $W$ boson from the observation of just
one charged lepton is generally impossible.

It can be shown that the ambiguity in the reconstruction of $y$ reduces
to the uncertainty in the knowledge of the transverse momentum $q_{T}$
of the $W$ boson \cite{RHICOverview}. If $q_{T}$ were known exactly,
the rapidity $y$ of the $W$ boson can be derived (up to a two-fold
ambiguity) from the measured rapidity $y_{\ell }$ and transverse
momentum $p_{T\ell }$ of the charged lepton (see the accompanying
paper \cite{PolWPhenomenology} for more details on this reconstruction
method). For instance, if all $W$ bosons were produced through the
Born process $q\bar{q}\rightarrow W$, the transverse momentum $q_{T}$
would be zero, and the correct solution for $y$ can be chosen statistically
for the leptons with large absolute values of the lepton rapidity
$y_{\ell }$, i.e., escaping near the beam pipe direction.

In reality, the $W^{\pm }$ bosons carry a non-zero transverse momentum
due to QCD radiation, with the most probable magnitude in the unpolarized
case of about 2 GeV. Most of the $W^{\pm }$ bosons obtain a non-zero
$q_{T}$ through the radiation of soft and collinear partons, which
cannot be approximated by finite-order perturbative calculations.
Since the power series in the strong coupling $\alpha _{S}$ does
not converge in the small-$q_{T}$ region, summation of dominant logarithmic
terms through all orders of this series is needed. This all-order
summation can be realized with the help of the methods that were studied
in a substantial detail \cite{DDT78,ParisiPetronzio79,AltarelliEllisMartinelli79,CollinsSoper81,CollinsSoper82,CSS,DWS,Arnold:1991yk,CPCsaba,Ladinsky:1994zn,Ellis:1997sc,Ellis:1998ii,Landry:1999an,Qiu:2000ga,Qiu:2000hf,Kulesza:2001jc,Landry:2002ix}
in the unpolarized case. Note, however, that the properties of the
multiple parton radiation depend on the spin of the initial hadrons.
The spin dependence of the collinear radiation off the external parton
lines can be seen from an explicit calculation. An additional spin
dependence can be contributed by the unknown nonperturbative dynamics
of strong interactions at large distances. Thus, conclusions about
the nature of the multiple parton radiation in the polarized hadronic
collisions cannot be inferred from the unpolarized case, and an additional
study is required to estimate the spin dependence of such radiation
under the RHIC conditions. 

The main goal of this paper is to provide a theoretical framework
for such a study. We present a complete formalism for resummation
in VBP with the proton beams of an arbitrary longitudinal polarization.
Furthermore, we explicitly account for the decay of the vector bosons
into the observed leptons, which is important because of the complicated
geometry of RHIC detectors. The results are derived at a one-loop
level of PQCD, and the summation of large logarithms is performed
with the help of the impact parameter resummation technique \cite{CSS}.
The detailed discussion of this lepton-level resummation formalism
in the unpolarized case can be found in Ref.~\cite{CPCsaba}. The
theoretical results are presented for arbitrary couplings of the vector
boson, which makes the results of this paper also applicable to lepton
pair production mediated by photons or $Z^{0}$ bosons. Our eventual
product is a numerical program for Monte-Carlo integration of fully
differential resummed cross sections in the presence of experimental
cuts. In the accompanying paper \cite{PolWPhenomenology}, we present
numerical results for the single-spin asymmetries in $W$ boson production
and estimate the sensitivity of these asymmetries to different models
of polarized PDFs. 

To review the already available literature, we note that the finite-order
asymmetries of order ${\mathcal{O}}(\alpha _{S})$ in the Drell-Yan
process are currently available for distributions in full lepton pair
momentum \cite{Weber1}, invariant mass \cite{Kamal1}, and rapidity
\cite{Gehrmann1}. Analogous distributions for $W$ boson production
were published in Refs.~\cite{Kamal2,Gehrmann2,Weber2}. Recently,
the ${\mathcal{O}}(\alpha _{S}^{2})$ fully differential distributions
of the Drell-Yan pairs were obtained in Ref.~\cite{Ravindran:2002na}.
Furthermore, Refs.$\, $\cite{Weber1,Weber2} presented the ${\mathcal{O}}(\alpha _{S})$
resummed single- and double-spin cross sections in the narrow width
approximation, and the ${\mathcal{O}}(\alpha _{S})$ Sudakov factor
was explicitly demonstrated to be independent from the polarization
of the proton beams. 
Note that most of the above publications 
(with the exception of the forward-backward asymmetry in Ref.~\cite{Kamal2} 
and polar angle distribution of the Drell-Yan 
leptons in the lab frame in Ref.~\cite{Kodaira:2003tq})
discuss the behavior of the whole lepton pair,  
rather than the distributions of the individual leptons.

In this paper, we go beyond the previous results in several aspects.
First, we present the  fully
differential finite-order and resummed 
cross sections for the decaying electroweak 
vector bosons, i.e.\emph{,}
we completely account for the spin correlations between the hadronic subsystem
and leptonic final state. In other words, the cross sections in this
work are fully differential in the individual lepton momenta, rather
than in the momentum of the lepton pair. These cross sections
are implemented in a Monte-Carlo integration program. Such lepton-level
analysis requires calculation of several additional angular structure
functions, which do not contribute in the narrow width approximation.
Furthermore, the $q_{T}$ resummation is needed not only for the parity-conserving
angular function $1+\cos ^{2}\theta $, which contributes to the boson-level
cross section, but also for the parity-violating angular function
$2\cos \theta $, which affects the angular distributions of the decay
products. The calculation for the parity-violating term $2\cos \theta $
is substantially more complex, because it involves a large number
of $\gamma _{5}$ matrices (and Levi-Civita tensors) coming from both
spin projection operators and axial couplings of the electroweak Lagrangian.
In contrast, the parity-conserving term $1+\cos ^{2}\theta $ depends
on $\gamma _{5}$ matrices and Levi-Civita tensors from the spin projection
operators only. 

It is well known that special care is needed to deal with the $\gamma _{5}$
matrices in $n\neq 4$ dimensions. In order to tackle this issue efficiently,
we have evaluated the $n$-dimensional cross sections using the dimensional
reduction method \cite{Siegel:1979wq,Schuler:1987ej,Korner:1994pv}.
We then converted our results to the conventional $\overline{MS}$
factorization scheme, which utilizes dimensional regularization \cite{'tHooft:1972fi,Breitenlohner:1977hr}.
From our calculation, we have found that the spin-dependent resummed
cross-sections in \cite{Weber1,Weber2} could not be used with the
existing parametrizations of the polarized PDFs in the $\overline{MS}$
scheme, due to the different factorization scheme used in those papers.
In addition, Ref.$\, $\cite{Weber2} has used an unconventional normalization
for the $n$-dimensional single-spin cross section in the $qG$ channel.
The present work has paid a special attention to rectify those inconsistencies
and obtain the hard cross sections compatible with the existing phenomenological
PDF parametrizations in the $\overline{MS}$ scheme.

The paper has the following structure. In Section~\ref{sec:Notations},
we introduce the notations for the kinematical variables and spin-dependent
cross sections. Section~\ref{sec:DREGandDRED} discusses the regularization
of cross section singularities by continuation of observables to $n\neq 4$
dimensions, as well as  the treatment of the $\gamma _{5}$ matrices.
In Section~ \ref{sec:Xsec}, we present in detail the ${\mathcal{O}}(\alpha _{S})$
finite-order and resummed cross sections. Section~\ref{sec:Summary}
contains the summary of our results and conclusions. The appendix
contains the expressions for all structure functions that contribute
to the lepton-level ${\mathcal{O}}(\alpha _{S})$ cross sections.

\section{\label{sec:Notations}Notations}

\subsection{Kinematical variables}

\begin{figure}
\begin{center}\includegraphics[  height=7cm]{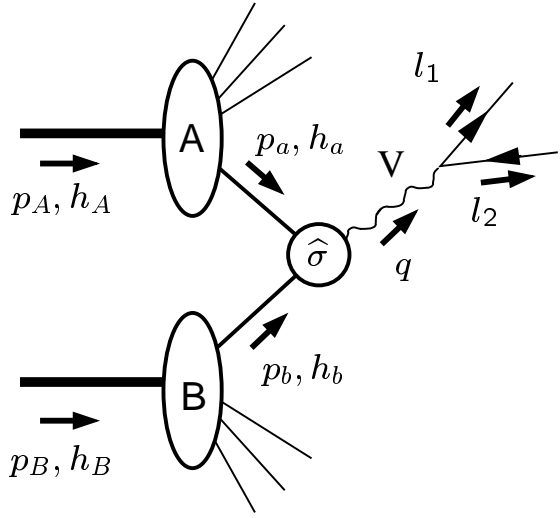}\end{center}

\caption{\label{fig:Momenta}Notations for the momenta and helicities of the
participating particles.}
\end{figure}

We discuss inclusive production of a lepton pair $A(p_{A})+B(p_{B})\rightarrow V^{*}(q)+X\rightarrow \ell _{1}(l_{1})+\bar{\ell }_{2}(l_{2})+X$
mediated by a virtual vector boson $V$. Our notations for the particle
momenta and helicities in this process are shown in Fig.~\ref{fig:Momenta}. 

The momentum $q^{\mu }$ of $V$ can be conveniently parametrized
by the invariant mass $Q$, rapidity $y$, and transverse momentum
$q_{T}$ of the vector boson in the laboratory frame. We have\begin{eqnarray}
Q^{2} & \equiv  & q^{\mu }q_{\mu };\\
y & \equiv  & \frac{1}{2}\ln \frac{q_{+}}{q_{-}};\\
q_{T}^{2} & \equiv  & -q_{t}^{\mu }q_{t\mu },
\end{eqnarray}
where $q_{t}^{\mu }$ is the component of the momentum $q^{\mu }$
that is orthogonal (in the relativistic sense) to the momenta $p_{A}^{\mu }$
and $p_{B}^{\mu }$:\begin{equation}
q_{t}^{\mu }\equiv q^{\mu }-\frac{(p_{A}\cdot q)}{(p_{A}\cdot p_{B})}p_{B}^{\mu }-\frac{(p_{B}\cdot q)}{(p_{A}\cdot p_{B})}p_{A}^{\mu }.\end{equation}
For any vector $n^{\mu }$, the {}``plus'' and {}``minus'' components
are defined as $n_{\pm }\equiv \left(n^{0}\pm n^{3}\right)/\sqrt{2}$. 

It will be useful to introduce the hadronic and partonic Mandelstam
invariants, defined by \begin{eqnarray}
s & = & (p_{A}+p_{B})^{2};\\
t & = & (p_{A}-q)^{2};\\
u & = & (p_{B}-q)^{2};\\
\shat  & = & (p_{a}+p_{b})^{2};\\
\that  & = & (p_{a}-q)^{2};\\
\uhat  & = & (p_{b}-q)^{2}.
\end{eqnarray}
A caret denotes quantities at the parton level. They are defined in
terms of the parton momenta $p_{a}^{\mu }\equiv \xi _{a}p_{A}^{\mu },$
$p_{b}^{\mu }\equiv \xi _{b}p_{B}^{\mu },$ where $\xi _{a}$ and
$\xi _{b}$ are the light-cone momentum fractions, which satisfy the
constraints $x_{A}\leq \xi _{a}\leq 1,$ $x_{B}\leq \xi _{b}\leq 1$,
with $x_{A,B}\equiv Qe^{\pm y}/\sqrt{s}$. Throughout the discussion,
all hadron masses are neglected. 

Our discussion will use two coordinate frames. The first frame is
the laboratory frame, or the center-of-mass frame of the initial proton
beams. The second frame is a special rest frame of the produced vector
boson (Collins-Soper frame \cite{CSframe}), in which the axis $Oz$
bisects the angle between the vectors $\vec{p}_{A}$ and $-\vec{p}_{B}$,
where $\vec{p_{A}}$ are $\vec{p_{B}}$ are the initial hadrons' momenta 
in this frame.
The coordinate transformation from the lab frame to the Collins-Soper
frame involves a boost in the direction of motion of $V$ and a rotation
around the $Oy$ axis (see Appendix A in \cite{CPCsaba} for the explicit
coordinate transformation matrix). The components of the momenta $p_{A}^{\mu }$
and $p_{B}^{\mu }$ in the Collins-Soper frame are\begin{equation}
p_{A,B}^{\mu }=\sqrt{\frac{s}{2}}\left(\frac{q_{\mp }}{Q},-\frac{q_{\mp }q_{T}}{M_{T}Q},0,\pm \frac{q_{\mp }}{M_{T}}\right),\end{equation}
where $M_{T}$ is the transverse mass of the vector boson, $M_{T}\equiv \sqrt{Q^{2}+q_{T}^{2}}.$
The momenta $l_{1}^{\mu }$ and $l_{2}^{\mu }$ of the final-state
leptons in this frame are\begin{eqnarray}
l_{1}^{\mu } & = & \frac{Q}{2}\Biggl (1,\sin \theta \cos \varphi ,\sin \theta \sin \varphi ,\cos \theta \Biggr ),\label{l1mu}\\
l_{2}^{\mu } & = & \frac{Q}{2}\Biggl (1,-\sin \theta \cos \varphi ,-\sin \theta \sin \varphi ,-\cos \theta \Biggr ).\label{l2mu}
\end{eqnarray}

Instead of using $l_{1}^{\mu }$ and $l_{2}^{\mu }$ directly, it
is convenient to operate with their linear combinations, \begin{equation}
q^{\mu }=l_{1}^{\mu }+l_{2}^{\mu }=\Biggl (Q,0,0,0\Biggr )\end{equation}
and\begin{equation}
l_{12}^{\mu }\equiv l_{1}^{\mu }-l_{2}^{\mu }=Q\Biggl (0,\sin \theta \cos \varphi ,\sin \theta \sin \varphi ,\cos \theta \Biggr ).\end{equation}
While $q^{\mu }$ describes the motion of the lepton pair as a whole,
$l_{12}^{\mu }$ specifies the motion of the individual leptons in
the decay of the vector boson. To separate the dynamics of vector
boson production from the dynamics of vector boson decay, we decompose
the lepton-level cross section in a sum over the functions $A_{\rho }(\theta ,\varphi )$
of the angles $\theta $ and $\varphi $ in the Collins-Soper frame:\begin{equation}
\frac{d\sigma }{dQ^{2}dydq_{T}^{2}d\Omega }=\sum _{\rho =-1}^{4}{}^{\rho }T(Q^{2},y,q_{T}^{2})A_{\rho }(\theta ,\varphi ).\end{equation}
In this equation, $d\Omega \equiv d\cos \theta \, d\varphi $ denotes
an element of the solid angle in the Collins-Soper frame. $^{\rho }T(Q^{2},y,q_{T}^{2})$
is the structure function corresponding to the angular function $A_{\rho }(\theta ,\varphi )$.
For an arbitrary chirality of the electroweak couplings, VBP at ${\mathcal{O}}(\alpha _{S})$
receives contributions from six angular functions $A_{\rho }(\theta ,\varphi )$:%
\footnote{At ${\mathcal{O}}(\alpha _{S}^{2})$, there will be additional contributions
to the longitudinally polarized cross section, which are proportional
to $\sin 2\theta \sin \varphi $ and $\sin ^{2}2\theta \sin \varphi $
\cite{Pire:1983tv}. These contributions lead to non-vanishing single-spin
asymmetries at $q_{T}\neq 0$ in the lepton distributions in the parity-conserving
case \cite{Pire:1983tv,Carlitz:1992fv,Nadolsky:1994uq}. Such contributions,
however, do not appear at the order ${\mathcal{O}}(\alpha _{S})$
discussed here.%
}\begin{eqnarray}
A_{-1}\equiv 1+\cos ^{2}\theta ; & \: \:  & A_{0}\equiv \frac{1}{2}(1-3\cos ^{2}\theta );\nonumber \\
A_{1}\equiv \sin 2\theta \cos \varphi ; & \: \:  & A_{2}\equiv \frac{1}{2}\sin ^{2}\theta \cos 2\varphi ;\nonumber \\
A_{3}\equiv 2\cos \theta ; & \: \:  & A_{4}\equiv \sin \theta \cos \varphi .\label{AngFns}
\end{eqnarray}
 The angular function $A_{-1}(\theta ,\varphi )$ is also sometimes
denoted as ${L}_{0}(\theta ,\varphi )$ (see, for instance, Ref.~\cite{CPCsaba}).
It will be shown below that the structure functions ${}^{-1}T,$ ${}^{0}T,$
${}^{1}T,$ and ${}^{2}T$ are generated by the vector part of the
electroweak current. On the other hand, the structure functions ${}^{3}T$
and ${}^{4}T$ are generated by the axial part of the electroweak
current. This feature leads to the following important consequence.
When the cross section is integrated over the complete solid angle
$\Omega $ of the lepton decay (as, e.g., in the calculation of a
boson-level cross section), all angular functions except $A_{-1}$
are integrated out. The resulting cross section is sensitive only
to the vector part of the electroweak current, so that, for instance,
the $W$ boson cross section can be straightforwardly derived from
the Drell-Yan pair cross section. On the other hand, if one is interested
in the angular distributions of the final-state leptons, the axial
part of the electroweak current cannot be ignored. In that case, additional
care is needed in the calculation of the parity-violating structure
functions ${}^{3}T$ and ${}^{4}T,$ which are affected by the $\gamma _{5}$
matrices from both the spin projection operators and electroweak couplings.

\subsection{Polarized cross sections}

We now introduce the special notations that will allow us to write
the cross sections for arbitrary proton polarizations in a compact
form. Let the polarized protons $A$ and $B$ have helicities $h_{A}$
and $h_{B}$, respectively. The allowed values are $+1$ for the right-handed
helicities and $-1$ for the left-handed helicities. The cross section
for this combination of the helicities is denoted as $\sigma ^{h_{A}h_{B}}$.
For brevity, the superscripts of $\sigma ^{h_{A}h_{B}}$ will only
show the signs of the helicities; that is, $\sigma ^{+1,+1}\equiv \sigma ^{++},$
$\sigma ^{+1,h_{B}}\equiv \sigma ^{+h_{B}},$ etc. 

The following combinations of the helicity cross sections will be
called {}``the cross sections that are unpolarized (U) or polarized
(P) on the side of the proton $A$'': \begin{eqnarray}
\sigma ^{(U,h_{B})} & \equiv  & \sigma ^{+h_{B}}+\sigma ^{-h_{B}},\label{sigmaUA}\\
\sigma ^{(P,h_{B})} & \equiv  & \sigma ^{+h_{B}}-\sigma ^{-h_{B}}.\label{sigmaPA}
\end{eqnarray}
Similarly, {}``the cross sections that are unpolarized (polarized)
on the side of the of the proton $B$'' are\begin{eqnarray}
\sigma ^{(h_{A},U)} & \equiv  & \sigma ^{h_{A}+}+\sigma ^{h_{A}-},\label{sigmaUB}
\end{eqnarray}
and\begin{eqnarray}
\sigma ^{(h_{A},P)} & \equiv  & \sigma ^{h_{A}+}-\sigma ^{h_{A}-},\label{sigmaPB}
\end{eqnarray}
respectively. 

Next, we introduce the unpolarized, single-spin, and double-spin cross
sections $\sigma ^{(U,U)},$ $\sigma ^{(P,U)},$ and $\sigma ^{(P,P)}$:\begin{eqnarray}
\sigma ^{(U,U)} & \equiv  & \frac{1}{4}\sum _{h_{A},h_{B}=\pm 1}\sigma ^{h_{A}h_{B}};\label{sigmaUU}\\
\sigma ^{(P,U)} & \equiv  & \frac{1}{4}\sum _{h_{A},h_{B}=\pm 1}h_{A}\, \sigma ^{h_{A}h_{B}};\label{sigmaPU}\\
\sigma ^{(P,P)} & \equiv  & \frac{1}{4}\sum _{h_{A},h_{B}=\pm 1}h_{A}h_{B}\, \sigma ^{h_{A}h_{B}}.\label{sigmaPP}
\end{eqnarray}
 The normalization factor 1/4 in (\ref{sigmaUU}-\ref{sigmaPP}) corresponds
to the number of the allowed helicity combinations for the initial-state
massless hadrons. The single-spin cross-section $\sigma ^{(P,U)}$
in Eq.~(\ref{sigmaPU}) corresponds to the polarized beam $A$ and
unpolarized beam $B$. The fourth independent linear combination of
the helicity cross-sections, the single-spin cross-section $\sigma ^{(U,P)}$,
can be obtained from the single-spin cross-section $\sigma ^{(P,U)}$
by interchanging the indices of the proton beams, $A\leftrightarrow B$.
At ${\mathcal{O}}(\alpha _{S})$, the single-spin cross section (\ref{sigmaPU})
is non-zero only if VBP violates parity with respect to the spatial
reflection. In other popular (but less uniform) notations, $\sigma ^{(U,U)},$
$\sigma ^{(P,U)}$, and $\sigma ^{(P,P)}$ are denoted as $\sigma ,$
$\Delta _{L}\sigma $, and $\Delta _{LL}\sigma $, respectively. Hence,
the conventional single- and double-spin asymmetries are constructed
as \begin{eqnarray}
A_{L} & = & \frac{\sigma ^{(P,U)}}{\sigma ^{(U,U)}},\\
A_{LL} & = & \frac{\sigma ^{(P,P)}}{\sigma ^{(U,U)}},
\end{eqnarray}
where the definition for the single-spin asymmetry $A_{L}$ does not
include an additional minus sign. This choice leads to the negative
values of $A_{L}$ for the left-handed interactions of fermions at
the Born level, and it is the same as the definition in Refs.~\cite{Gehrmann2,Kamal2}. 

The QCD factorization expresses the hadron-level cross sections $\sigma _{AB}^{h_{A}h_{B}}$
in terms of the {}``hard'' parton-level cross sections $\widehat{\sigma }_{ab}^{h_{a}h_{b}}(\mu _{F})$
and parton distribution functions: \begin{equation}
\sigma _{AB}^{h_{A}h_{B}}=\sum _{a,b=G,\stackrel{(-)}{u},\stackrel{(-)}{d},\dots }\sum _{h_{a},h_{b}=\pm 1}\int _{0}^{1}d\xi _{a}\int _{0}^{1}d\xi _{b}\widehat{\sigma }_{ab}^{h_{a}h_{b}}(\mu _{F})f_{h_{a}/h_{A}}(\xi _{a},\mu _{F})f_{h_{b}/h_{B}}(\xi _{b},\mu _{F}).\label{fact}\end{equation}
Here $f_{h_{a}/h_{A}}(\xi _{a},\mu _{F})$ denotes a helicity-dependent
parton distribution function (PDF), \textit{\emph{i.e.}}, the probability
of finding a parton $a$ with the momentum $p_{a}^{\mu }=\xi _{a}p_{A}^{\mu }$
and helicity $h_{a}$ in a hadron $A$ with the momentum $p_{A}^{\mu }$
and helicity $h_{A}.$ The parton-level cross section $\widehat{\sigma }_{ab}^{h_{a}h_{b}}(\mu _{F})$
is separated from the PDFs $f_{h_{a}/h_{A}}(\xi _{a},\mu _{F})$ at
a factorization scale $\mu _{F}$, which in our calculation is assumed
to coincide with the QCD renormalization scale.

Since the strong interactions conserve parity, only two linear combinations
of the helicity-dependent PDFs $f_{h_{a}/h_{A}}(\xi _{a},\mu _{F})$
are independent. These combinations will be called the unpolarized
and polarized PDFs, respectively:\begin{eqnarray}
f_{a/A}^{(U)}(\xi _{a},\mu _{F}) & \equiv  & f_{+/+}(\xi _{a},\mu _{F})+f_{-/+}(\xi _{a},\mu _{F})=f_{+/-}(\xi _{a},\mu _{F})+f_{-/-}(\xi _{a},\mu _{F}),\label{fU}\\
f_{a/A}^{(P)}(\xi _{a},\mu _{F}) & \equiv  & f_{+/+}(\xi _{a},\mu _{F})-f_{-/+}(\xi _{a},\mu _{F})=-f_{+/-}(\xi _{a},\mu _{F})+f_{-/-}(\xi _{a},\mu _{F}).\label{fP}
\end{eqnarray}
In common notations, $f_{a/A}^{(U)}(\xi _{a},\mu _{F})\equiv f_{a/A}(\xi _{a},\mu _{F})$,
and $f_{a/A}^{(P)}(\xi _{a},\mu _{F})\equiv \Delta f_{a/A}(\xi _{a},\mu _{F}).$
When written in terms of the unpolarized and polarized cross sections,
the factorization formula (\ref{fact}) sums only over the types of
the partons and not over their helicities: \begin{eqnarray}
\sigma _{AB}^{(H_{A},H_{B})} & = & \sum _{a,b=G,\stackrel{{(-)}}{u},\stackrel{{(-)}}{d},...}\int _{0}^{1}d\xi _{a}\int _{0}^{1}d\xi _{b}\widehat{\sigma }_{ab}^{(H_{A},H_{B})}(\mu _{F})f_{a/A}^{(H_{A})}(\xi _{a},\mu _{F})f_{b/B}^{(H_{B})}(\xi _{b},\mu _{F}),\label{sigmaHAHB}
\end{eqnarray}
where $H_{A},H_{B}$ denote unpolarized or polarized quantities:\[
H_{A,B}\equiv U,\, P.\]
Eq.~(\ref{sigmaHAHB}) already illustrates the usefulness of the
indices $U$ and $P$: in one equation, it covers the factorized representations
for all three cases of the unpolarized, single-spin, and double-spin
cross sections. In the following parts of the paper, many expressions
will be presented in this compact and uniform notation.

\subsection{Constant overall factors}

\begin{table}

\caption{\label{table:Couplings}The couplings of the electroweak vector bosons
$V$ to the fermions. The left couplings are $f_{L},g_{L}$, and the
right couplings are $f_{R},g_{R}$; see Eqs.~(\ref{fLR}) and (\ref{gLR})
for their definition. $g$ is the electroweak coupling, $\alpha _{EM}(\mu )$
is the running fine structure constant, and $s_{w}\equiv \sin \theta _{w}$
($c_{w}\equiv \cos \theta _{w}$) is the sine (cosine) of the weak
mixing angle. $Q_{f}$ is the fractional charge of the fermions ($Q_{u}=2/3$,
$Q_{d}=-1/3$, $Q_{\nu }=0$, $Q_{e^{-}}=-1$). $T_{3}$ is the eigenvalue
of the weak isospin for the fermion ($T_{3}^{u}=T_{3}^{\nu }=1/2$,
$T_{3}^{d}=T_{3}^{e^{-}}=-1/2$). }

\begin{center}\begin{tabular}{ccc}
\hline 
$V$&
$f_{L},\, g_{L}$&
$f_{R},\, g_{R}$\\
\hline 
$\gamma $&
$\frac{Q_{f}}{2}\sqrt{4\pi \alpha _{EM}(\mu )}$&
$\frac{Q_{f}}{2}\sqrt{4\pi \alpha _{EM}(\mu )}$\\
$W^{\pm }$&
$g/(2\sqrt{2})$&
0\\
$Z^{0}$&
~~~$g(T_{3}-Q_{f}s_{w}^{2})/(2c_{w})$~~~&
$-gQ_{f}s_{w}^{2}/(2c_{w})$\\
\hline
\end{tabular}\end{center}
\end{table}

The results in this paper cover production of virtual photons, $W$,
and $Z$ bosons. We use the on-shell scheme and improved Born approximation
to parametrize the electroweak parameters in the considered processes.
Let $f_{L}$, $f_{R}$ and $g_{L}$, $g_{R}$ be the chiral couplings
entering the leptonic vertex $\ell _{1}\bar{\ell }_{2}V$ and quark
vertex $q_{j}\bar{q}_{\bar{k}}V$, respectively. Here the quark flavor
indices are $j=u,d,s,...$ and $\bar{k}=\bar{u},\bar{d},\bar{s},...$
.The corresponding interaction Lagrangians are expressed as\begin{equation}
{\mathcal{L}}_{\ell _{1}\bar{\ell }_{2}V}=i\gamma _{\mu }\Biggl (f_{L}(1-\gamma _{5})+f_{R}(1+\gamma _{5})\Biggr ),\label{fLR}\end{equation}
and\begin{equation}
{\mathcal{L}}_{q_{j}\bar{q}_{\bar{k}}V}=i\gamma _{\mu }\Biggl (g_{L}(1-\gamma _{5})+g_{R}(1+\gamma _{5})\Biggr )V_{j\bar{k}},\, \label{gLR}\end{equation}
respectively. The matrix $V_{j\bar{k}}$ defines the flavor structure
of the $q_{j}\bar{q}_{\bar{k}}V$ vertex and is given by\begin{equation}
V_{j\bar{k}}=V_{\bar{k}j}=\left\{ \begin{array}{c}
 \mbox {CKM\quad matrix\quad elements\quad for\quad }W^{\pm },\\
 \delta _{jk}\mbox {\quad for\quad }\gamma ^{*},Z^{0}.\end{array}
\right..\label{Vij}\end{equation}
The values of the leptonic chiral couplings $f_{L}$, $f_{R}$ and
quark chiral couplings $g_{L}$, $g_{R}$ to $\gamma ^{*},$ $W^{\pm }$,
and $Z^{0}$ bosons are specified in Table~\ref{table:Couplings}.
They are expressed in terms of the running fine structure constant
$\alpha _{EM}(\mu ),$ weak coupling $g,$ and sine of the weak angle
$\sin \theta _{W}$, calculated as\begin{equation}
g^{2}=4\sqrt{2}M_{W}^{2}G_{F}\end{equation}
and\begin{equation}
\sin ^{2}\theta _{W}=1-M_{W}^{2}/M_{Z}^{2}\end{equation}
from the input values of the Fermi constant $G_{F}$, $W$ boson mass
$M_{W}$, and $Z^{0}$ boson mass $M_{Z}$. The nice feature about
this parametrization of the electroweak couplings is that the higher-order
electroweak radiative corrections in VBP are reduced.

Various constant factors in front of the cross sections will be absorbed
in the hadronic normalization constants $\sigma _{ab}^{\pm }$ and
leptonic normalization constants $\sigma _{\ell }^{\pm }$: \begin{eqnarray}
\sigma _{ab}^{\pm } & \equiv  & \frac{2\pi }{N_{c}}\left(g_{L}^{2}\pm g_{R}^{2}\right)|V_{ab}|^{2},\quad a,b=G,\, u,\, \bar{u},\, d,\, \bar{d},\dots \quad ;\label{sigmaab}\\
\sigma _{\ell }^{\pm } & \equiv  & \frac{1}{32\pi ^{2}}\frac{Q^{2}}{(Q^{2}-M_{V}^{2})^{2}+\Gamma _{V}^{2}Q^{4}/M_{V}^{2}}\left(f_{L}^{2}\pm f_{R}^{2}\right).\label{sigmal}
\end{eqnarray}
In these equations, $N_{c}=3$ is the number of colors, $M_{V}$ and
$\Gamma _{V}$ denote the mass and width of the vector boson. The
indices $a$ and $b$ denote all possible active parton flavors, including
the gluons. When both $a$ and $b$ correspond to the quarks or antiquarks
($a,b=u,\, \bar{u},\, d,\, \bar{d},\dots $), the matrix element $V_{ab}$
is given by Eq.~(\ref{Vij}). If one of the indices in $V_{ab}$
corresponds to the gluons, we implicitly assume summation over the
quarks or antiquarks on the side of the gluon:\begin{equation}
|V_{jG}|^{2}=\sum _{\bar{k}=\bar{u},\bar{d},\dots }|V_{j\bar{k}}|^{2},\quad |V_{G\bar{k}}|^{2}=\sum _{j=u,d,\dots }|V_{j\bar{k}}|^{2},\mbox {\quad etc.}\end{equation}
 Unless explicitly stated otherwise, the subscripts $j$ and $\bar{k}$
run over the quark and antiquark flavors, respectively ($j=u,d,...$;
and $\bar{k}=\bar{u},\bar{d},...)$. They should be distinguished
from the indices $a$ and $b$, which can take the value of any parton
flavor ($a,b=G,u,\bar{u},d,\bar{d},...$).

\section{Regularization of singularities: general procedure\label{sec:DREGandDRED}}

\begin{figure}
\begin{center}\includegraphics[  width=0.80\textwidth]{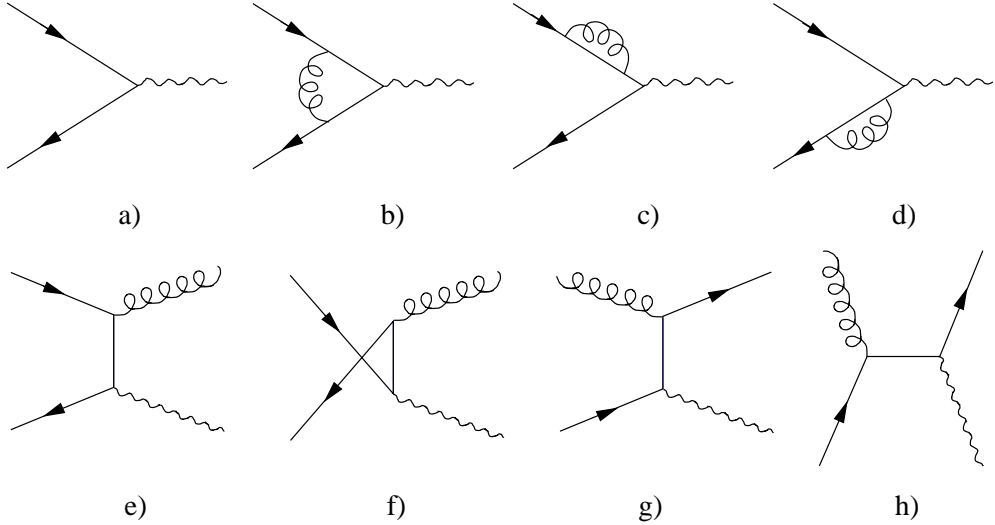}\end{center}

\caption{Partonic subprocesses contributing to vector boson production.\label{fig:FeynmanDiagrams}}
\end{figure}

The complete set of Feynman diagrams contributing at order ${\mathcal{O}}(\alpha _{S})$
is shown in Fig.~\ref{fig:FeynmanDiagrams}. The calculation involves
cancellation of ultraviolet singularities in the sum of all virtual
diagrams (Figs.~\ref{fig:FeynmanDiagrams}b-\ref{fig:FeynmanDiagrams}d),
cancellation of soft singularities in the sum of virtual and real
emission corrections (Figs.~\ref{fig:FeynmanDiagrams}b-\ref{fig:FeynmanDiagrams}h),
and factorization of collinear singularities in real emission corrections
(Figs.~\ref{fig:FeynmanDiagrams}e-\ref{fig:FeynmanDiagrams}h).
Those singularities should be first exposed by intermediate regularization,
which can be achieved by continuation to $n=4-2\eps $ dimensions.
In our case, this approach involves the necessity to define $\gamma _{5}$
matrices in $n\neq 4$ dimensions. As is well known, such continuation
cannot be mathematically consistent and preserve symmetries of the
classical Lagrangian at the same time. For instance, dimensional regularization
(DREG) \cite{'tHooft:1972fi,Bollini:1972ui,Bollini:1972bi} assumes
that the $\gamma _{5}$ matrix in $n>4$ dimensions is a purely four-dimensional
object satisfying the following commutation relations:\[
\{\gamma _{5},\gamma _{\mu }\}=0,\, \mbox {if\, }\mu \leq 4;\, [\gamma _{5},\gamma _{\mu }]=0,\mbox {\, if\, }\mu >4.\]
Since the conventional $\overline{MS}$ scheme utilizes dimensional
regularization, we will sometimes refer to this scheme also as the
{}``DREG factorization scheme''. The DREG choice is mathematically
consistent \cite{Breitenlohner:1977hr}. At the same time, it requires
to treat vector components in the $(n-4)$-dimensional subspace differently
from vector components in the four-dimensional subspace. The proliferation
of extra $(n-4)$-dimensional terms considerably complicates the algebra
and, more importantly, leads to the violation of chiral symmetry \cite{CollinsRenormalizationBook}
and supersymmetry \cite{Capper:1980ns,Avdeev:1980bh}. In spin-dependent
QCD, this implies non-conservation of the quark helicity in the process
of gluon radiation. Those symmetries have to be restored order by
order by introducing additional counterterms. 
At order ${\cal O}(\alpha_S)$, 
a well-known example of this feature is provided by an additional 
renormalization of the one-loop virtual QCD corrections needed
to restore the Ward identity for the axial electroweak current
\cite{CollinsRenormalizationBook}. Another relevant example 
is given below in the discussion of 
Eq.~(\ref{PPqqeps}), where an additional finite renormalization
is performed in DREG in the part associated with the quark-gluon splitting 
in $n\neq 4$ dimensions. In both cases, the additional renormalization
restores conservation of helicity in the processes with radiation of soft or
collinear partons, which is otherwise violated 
by the evanescent $(n-4)$-dimensional terms appearing in the DREG calculation.

On the other hand, alternative approaches sacrifice certain features
of the full theory, such as the cyclic permutability in Dirac traces
\cite{Korner:1992sx} in the anticommuting $\gamma _{5}$ 
scheme \cite{Chanowitz:1978uj,Chanowitz:1979zu};
or they are mathematically inconsistent at higher orders of the perturbative
expansion. In a popular alternative to DREG, dimensional reduction
(DRED) \cite{Siegel:1979wq,Schuler:1987ej,Korner:1994pv}, the spinor
indices are kept in four dimensions, while particle momenta are declared
to have $n<4$ components. By first evaluating the Dirac traces in
four dimensions and then evaluating loop integrals in $n$ dimensions,
one explicitly preserves the quark helicity in the QCD vertex. Since
the DRED method treats all spin components on the same footing, it
is also algebraically simpler than the DREG approach. Nonetheless,
contradictions in the DRED framework are present in the diagrams with
more than two loops \cite{Siegel:1980qs,Avdeev:1983xy}. These contradictions,
which reflect inconsistency in the decomposition of the 4-dimensional
momentum space into $4$-dimensional subspace (associated with the
spin indices) and $(4-n)$-dimensional subspace (associated with the
momentum indices), do not occur at lower orders of the perturbative
expansion. Hence, it is justified to use the DRED method at the one-loop
order of PQCD.

Due to the large number of $\gamma _{5}$ matrices to be handled,
and to avoid entirely the issue of additional symmetry-enforcing 
renormalizations in the DREG scheme,
we have chosen to perform the current one-loop calculation in the
DRED framework. 
In particular, soft or collinear radiative corrections in 
DRED automatically preserve the chirality structure of the 
electroweak vertex, so that such corrections do not affect
the angular distributions of the leptons observed at the Born level 
(cf.\, further discussion in section \ref{sec:Xsec}).
Once the DRED results are available, they can be transformed
to the $\overline{MS}$ (DREG) factorization scheme to make them compatible
with the existing phenomenological PDF parametrizations. The detailed
description of this transformation can be found in 
Refs.~\cite{Schuler:1987ej,Kunszt:1994sd,Korner:1994pv,Kamal1,Kamal2},
and we refer the reader to those papers 
for more information.
In the following, we will outline only those aspects of this method that
are relevant to the derivation of the resummed cross sections. 

We start by calculating the parton-level cross sections\begin{equation}
\frac{d\sigma _{a'b'}^{(H_{a'},H_{b'})}}{dQ^{2}dydq_{T}^{2}d\Omega }=\frac{1}{4\shat }\frac{1}{(4\pi )^{4}}\delta \left[\shat +\that +\uhat -Q^{2}\right]H_{\mu \nu }L^{\mu \nu }.\label{SigmaabUnfactorized}\end{equation}
In this equation, the (purely four-dimensional) leptonic tensor $L^{\mu \nu }$
is given by\begin{eqnarray}
L^{\mu \nu } & \equiv  & 4\Biggl \{\left[f_{L}^{2}+f_{R}^{2}\right]\left(-q^{2}g^{\mu \nu }+q^{\mu }q^{\nu }-l_{12}^{\mu }l_{12}^{\nu }\right)+i\left[f_{L}^{2}-f_{R}^{2}\right]\epsilon ^{\mu \nu \alpha \beta }q_{\alpha }(l_{12})_{\beta }\Biggr \},
\end{eqnarray}
and the hadronic tensor $H_{\mu \nu }$ includes squared matrix elements
corresponding to the Feynman diagrams in Fig.~\ref{fig:FeynmanDiagrams}.
We first evaluate the Dirac traces with the help of the package TRACER
\cite{Jamin:1993dp} and keeping all particle spins in four dimensions.
The projections on the unpolarized ($U$) and polarized ($P$) initial
parton states are realized by inserting an appropriate spin projection
operator\begin{equation}
{\mathcal{P}}_{q}^{(U)}=\frac{1}{2}I,\mbox {\, or\, }{\mathcal{P}}_{q}^{(P)}=\frac{1}{2}\gamma _{5}\end{equation}
 on each incoming quark leg, and \begin{equation}
{\mathcal{P}}_{G}^{\alpha \beta (U)}=\frac{1}{2}\left\{ -g^{\alpha \beta }+2\frac{p_{a'}^{\alpha }p_{b'}^{\beta }+p_{a'}^{\beta }p_{b'}^{\alpha }}{\shat }\right\} ,\mbox {\, or\, }{\mathcal{P}}_{G}^{\alpha \beta (P)}=\frac{i}{\shat }\epsilon ^{\alpha \beta \gamma \delta }p_{a'\gamma }p_{b'\delta }\end{equation}
on each incoming gluon leg. 

We then contract the resulting four-dimensional tensors with $n$-dimensional
particle momenta, neglecting the mismatch between the four- and $n$-dimensional
indices. We find that the individual virtual diagrams (Figs.~\ref{fig:FeynmanDiagrams}b-\ref{fig:FeynmanDiagrams}d)
contain ultraviolet singularities, which, however, cancel in the sum
of all virtual diagrams. Furthermore, the real emission contributions
contain the singularities when the unobserved final-state parton is
soft or collinear to one of the incoming hadrons. In the soft and
collinear limits the transverse momentum $q_{T}$ approaches zero.
The soft singularities (appearing as the poles proportional to $1/\eps ^{2}$
and $1/\eps $) cancel in the sum of all ${\mathcal{O}}(\alpha _{S})$
diagrams. The remaining task is to factorize the collinear singularities
according to the factorization scheme that was used to define the
hadron-level PDFs.

It can be shown that the choice of the $\gamma_5$ prescription affects
only functions ${\mathcal{C}}_{a/b}^{(H)}(\xi ,b,\mu )$ (where $H=U$
or $P$) in the resummed cross section (see the detailed discussion
of these functions in the next section). The ${\mathcal{C}}$-functions
contain $(n-4)$-dimensional parts of the $n$-dimensional splitting
functions, which are different in the DRED and DREG schemes. An ${\mathcal{O}}(\alpha _{S})$
$n$-dimensional splitting function $P_{a/b}^{(H),n}(\xi )$ can be
calculated in the same way as the conventional four-dimensional splitting
function $P_{a/b}^{(H)}(\xi )$, except that it also retains the terms
of order ${\mathcal{O}}(\eps )$. Hence, it contains an additional
finite part $\eps P_{a/b}^{(H),\eps }(\xi )$, which vanishes as $n\rightarrow 4$:\begin{eqnarray}
P_{a/b}^{(H),n}(\xi ) & = & P_{a/b}^{(H)}(\xi )+\eps P_{a/b}^{(H),\eps }(\xi );\\
\lim _{n\rightarrow 4}P_{a/b}^{(H),n}(\xi ) & = & P_{a/b}^{(H)}(\xi ).
\end{eqnarray}

The four-dimensional splitting functions $P_{a/b}^{(H)}(\xi )$ needed
for our calculations are \begin{eqnarray}
P_{q/q}^{(U)}(\xi ) & = & =C_{F}\left(\frac{1+\xi ^{2}}{1-\xi }\right)_{+},\label{PUqq}\\
P_{q/G}^{(U)}(\xi ) & = & \frac{1}{2}\left(\xi ^{2}+(1-\xi )^{2}\right)
\end{eqnarray}
in the unpolarized case \cite{Dokshitzer77,GribovLipatov72,AP}, and\begin{eqnarray}
P_{q/q}^{(P)}(\xi ) & = & C_{F}\left(\frac{1+\xi ^{2}}{1-\xi }\right)_{+}=P_{q/q}^{(U)}(\xi ),\\
P_{q/G}^{(P)}(\xi ) & = & 2\xi -1\label{PPqG}
\end{eqnarray}
in the spin-dependent case \cite{AhmedRoss,AP}. In the unfactorized
parton-level cross section (\ref{SigmaabUnfactorized}), the functions
$P_{a/b}^{(H),n}(\xi )$, which are implicitly contained in the hadronic
tensor $H_{\mu \nu }$, are multiplied by the collinear poles $1/\eps $,
i.e., they appear in the terms\begin{equation}
-\frac{\alpha _{S}}{2\pi }\frac{1}{\eps }P_{a/b}^{(H),n}(\xi )=-\frac{\alpha _{S}}{2\pi }\frac{1}{\eps }P_{a/b}^{(H)}(\xi )-\frac{\alpha _{S}}{2\pi }P_{a/b}^{(H),\eps }(\xi ).\label{Pn}\end{equation}
The first piece on the right-hand side of Eq.~(\ref{Pn}) is an ${\mathcal{O}}(\alpha _{S})$
contribution to the parton-level distribution function $f_{a/b}^{(H)}(\xi ,\eps )$.
Hence, in accordance with the factorization theorem (\ref{sigmaHAHB})
for the case when $A=a',$ and $B=b'$, it is factorized out of the
parton-level cross section $\sigma _{a'b'}$ to obtain the hard section
$\wh{\sigma }_{a^{\prime \prime }b^{\prime \prime }}$. On the other
hand, the second piece in Eq.~(\ref{Pn}) remains in the hard part.
As a result, the terms $P_{a/b}^{(H),\eps }(\xi )$ are included in
the ${\mathcal{O}}(\alpha _{S}/\pi )$ coefficients ${\mathcal{C}}_{a/b}^{(H)[1]}(\xi ,b\mu )$
of the ${\mathcal{C}}$-functions, as

\begin{eqnarray}
\C _{i/j}^{(H)[1]}(\xi ,b\mu ) & = & \delta _{ij}\Biggl \{-\frac{1}{2}P_{q/q}^{(H),\eps }(\xi )-P_{q/q}^{(H)}(\xi )\ln \Bigl (\frac{\mu b}{b_{0}}\Bigr )\nonumber \\
 & - & C_{F}\delta (1-\xi )\biggl [\frac{19}{16}-\frac{\pi ^{2}}{4}+\ln ^{2}\Bigl (\frac{e^{-3/4}C_{1}}{b_{0}C_{2}}\Bigr )\biggr ]\Biggr \},\label{C1Hij}
\end{eqnarray}
and\begin{eqnarray}
\C _{i/G}^{(H)[1]}(\xi ,b\mu ) & = & -\frac{1}{2}P_{q/G}^{(H),\eps }(\xi )-P_{q/G}^{(H)}(\xi )\ln \Bigl (\frac{\mu b}{b_{0}}\Bigr );\, i,j=u,\bar{u},d,\bar{d},\dots \, .\label{C1Hig}
\end{eqnarray}
The meaning of the constants $C_{1},$ $C_{2}$, and $b_{0}$ is discussed
in the next section.

Since in the DRED scheme the particle spins are exactly four-dimensional,
the $(n-4)$-dimensional parts $P_{a/b}^{(H),\eps }(\xi )$ of the
DRED splitting functions identically vanish:\begin{equation}
\mbox {DRED:\, }P_{a/b}^{(H),\eps }(\xi )=0.\label{PHepsDRED}\end{equation}
 Therefore, in the DRED scheme the ${\mathcal{C}}$-functions (\ref{C1Hij})
and (\ref{C1Hig}) reduce to \begin{eqnarray}
\mbox {DRED:\, }\C _{i/j}^{(H)[1]}(\xi ,b\mu ) & = & \delta _{ij}\Biggl \{-P_{q/q}^{(H)}(\xi )\ln \Bigl (\frac{\mu b}{b_{0}}\Bigr )\nonumber \\
 & - & C_{F}\delta (1-\xi )\biggl [\frac{19}{16}-\frac{\pi ^{2}}{4}+\ln ^{2}\Bigl (\frac{e^{-3/4}C_{1}}{b_{0}C_{2}}\Bigr )\biggr ]\Biggr \},\label{C1HijDRED}\\
\mbox {\, \, \, \, \, }\C _{i/G}^{(H)[1]}(\xi ,b\mu ) & = & -P_{q/G}^{(H)}(\xi )\ln \Bigl (\frac{\mu b}{b_{0}}\Bigr );\, i,j=u,\bar{u},d,\bar{d},\dots \, .\label{C1HiGDRED}
\end{eqnarray}

In the DREG scheme, the $(n-4)$-dimensional parts of the unpolarized
\cite{Ellis:1981wv} and polarized \cite{Gordon:1993qc} splitting
functions are given by\begin{eqnarray}
\mbox {DREG:\, }P_{q/q}^{(U),\eps }(\xi ) & = & C_{F}\left[-(1-\xi )+\frac{1}{2}\delta (1-\xi )\right],\label{PHepsDREG1}\\
\mbox {\, \, \, \, \, }P_{q/G}^{(U),\eps }(\xi ) & = & \xi ^{2}-\xi ,\\
\mbox {\, \, \, \, \, }P_{q/q}^{(P),\eps }(\xi ) & = & C_{F}\left[3(1-\xi )+\frac{1}{2}\delta (1-\xi )+z_{\eps }(\xi )\right]\nonumber \\
 & = & C_{F}\left[-(1-\xi )+\frac{1}{2}\delta (1-\xi )\right]=P_{q/q}^{(U),\eps }(\xi ),\label{PPqqeps}\\
\mbox {\, \, \, \, \, }P_{q/G}^{(P),\eps }(\xi ) & = & \xi -1.\label{PHepsDREG4}
\end{eqnarray}
In the quark-initiated polarized splitting function $P_{q/q}^{(P),\eps }(\xi $)
(see Eq.~(\ref{PPqqeps})), an additional finite renormalization
term $z_{\eps }(\xi )=-4C_{F}(1-\xi )$ was included to restore the
quark helicity conservation in the quark-gluon vertex, in accordance
with the existing definition in the DREG scheme \cite{Mertig:1996ny,Vogelsang:1996vh,VogelsangTwoLoop}.
With the help of Eqs.~(\ref{C1Hij}) and (\ref{C1Hig}), the ${\mathcal{C}}$-functions
in the DREG scheme are found to be\begin{eqnarray}
\mbox {DREG:\, }\C _{i/j}^{(U)[1]}(\xi ,b\mu ) & = & \C _{i/j}^{(P)[1]}(\xi ,b\mu )=\nonumber \\
 & = & \delta _{ij}\Biggl \{\frac{C_{F}}{2}(1-\xi )-P_{q/q}^{(U)}(\xi )\ln \Bigl (\frac{\mu b}{b_{0}}\Bigr )\nonumber \\
 & - & C_{F}\delta (1-\xi )\biggl [\frac{23}{16}-\frac{\pi ^{2}}{4}+\ln ^{2}\Bigl (\frac{e^{-3/4}C_{1}}{b_{0}C_{2}}\Bigr )\biggr ]\Biggr \},\label{C1HijDREG}\\
\mbox {\, \, \, \, \, }\C _{i/G}^{(U)[1]}(\xi ,b\mu ) & = & \frac{1}{2}\xi (1-\xi )-P_{q/G}^{(U)}(\xi )\ln \Bigl (\frac{\mu b}{b_{0}}\Bigr ),\label{C1UiGDREG}
\end{eqnarray}
and\begin{eqnarray}
\mbox {\, \, \, \, \, }\C _{i/G}^{(P)[1]}(\xi ,b\mu ) & = & \frac{1}{2}(1-\xi )-P_{q/G}^{(P)}(\xi )\ln \Bigl (\frac{\mu b}{b_{0}}\Bigr ).\label{C1PiGDREG}
\end{eqnarray}

To summarize this part, the difference between the DRED and DREG factorization
schemes is contained entirely in the functions ${\mathcal{C}}_{a/b}^{(H)}(\xi ,b,\mu )$
of the resummed cross section. The form of the scheme-dependent terms
in the ${\mathcal{C}}$-functions is determined by the $n$-dimensional
splitting functions in each factorization scheme. Hence, to transform
the results from the DRED scheme to the DREG scheme, one simply finds
where the DRED $(n-4)$-dimensional functions $P_{a/b}^{(H),\eps }(\xi )$
would appear in the ${C}$-functions if they were not identically
zero. One then inserts at those places the DREG functions $P_{a/b}^{(H),\eps }(\xi )$
given in Eqs.~(\ref{PHepsDREG1})-(\ref{PHepsDREG4}).

\section{Cross sections \label{sec:Xsec}}

\subsection{Finite-order results}

Let us now explicitly present the VBP cross sections, starting from
the lowest order (LO) hadron-level cross section (Fig.~\ref{fig:FeynmanDiagrams}a):
\begin{eqnarray}
\left.\frac{d\sigma _{AB}^{(H_{A},H_{B})}}{dQ^{2}dyd^{2}\vec{q}_{T}d\Omega }\right|_{LO} & = & \frac{1}{s}\delta ^{2}\left(\vec{q}_{T}\right)\sum _{j=u,d,\dots }\, \sum _{\bar{k}=\bar{u},\bar{d},\dots }\Biggl [\GHAHB f_{j/A}^{(H_{A})}(x_{A},\mu _{F})f_{\bar{k}/B}^{(H_{B})}(x_{B},\mu _{F})\nonumber \\
 & + & (j\leftrightarrow \bar{k})\Biggr ].\label{LO}
\end{eqnarray}
As discussed in Section~\ref{sec:Notations}, $H_{A}$ and $H_{B}$
can take one of the two values ($U$ or $P$) corresponding to the
cross section that is unpolarized (or polarized) on the side of the
respective initial-state hadron. The index $j$ runs over the active
quark flavors, while $\bar{k}$ runs over the active antiquark flavors.
The functions $G_{j\bar{k}}^{(H_{A},H_{B})}(\theta ,\varphi )$ in
Eq.~(\ref{LO}) are composed of the normalization factors $\sigma _{l}^{\pm },\, \sigma _{j\bar{k}}^{\pm }$
from Eqs.~(\ref{sigmaab}), (\ref{sigmal}) and angular functions
$A_{-1}(\theta ,\varphi )$ and $A_{3}(\theta ,\varphi )$ from Eq.~(\ref{AngFns}):
\begin{eqnarray}
G_{j\bar{k}}^{(U,U)}(\theta ,\varphi ) & = & \sigma _{\ell }^{+}\sigma _{j\bar{k}}^{+}A_{-1}(\theta ,\varphi )+\epsilon _{j\bar{k}}\sigma _{\ell }^{-}\sigma _{j\bar{k}}^{-}A_{3}(\theta ,\varphi ),\nonumber \\
G_{j\bar{k}}^{(P,U)}(\theta ,\varphi ) & = & -\epsilon _{j\bar{k}}\sigma _{\ell }^{+}\sigma _{j\bar{k}}^{-}A_{-1}(\theta ,\varphi )-\sigma _{\ell }^{-}\sigma _{j\bar{k}}^{+}A_{3}(\theta ,\varphi ),\nonumber \\
G_{j\bar{k}}^{(P,P)}(\theta ,\varphi ) & = & -G_{j\bar{k}}^{(U,U)}(\theta ,\varphi ).\label{gees}
\end{eqnarray}
The flavor-space tensor $\epsilon _{j\bar{k}}$ determines the change
of the sign under the interchange of quarks and antiquarks:\begin{equation}
\epsilon _{j\bar{k}}=1,\quad \epsilon _{\bar{k}j}=-1.\end{equation}
Note that the parity-violating angular function ${\mathcal{A}}_{3}=\cos \theta $
does not contribute if $V$ couples to the fermions through the vector
current, i.e.\emph{,} if $\sigma _{\ell }^{-}=\sigma _{j\bar{k}}^{-}=0.$
From the last equation (\ref{gees}), we see that the double-spin
parton-level cross section is equal to the unpolarized cross section
with the minus sign. This equality is a direct consequence of the
chirality conservation in the electroweak couplings (\ref{fLR}) and
(\ref{gLR}).

Next, consider the ${\mathcal{O}}(\alpha _{S})$ virtual corrections
shown in Figs.~\ref{fig:FeynmanDiagrams}b-\ref{fig:FeynmanDiagrams}d.
In $n=4-2\varepsilon $ dimensions, the contribution of the virtual
corrections is given by the LO cross section scaled by a function
$\Phi (\alpha _{S},\varepsilon )$:

\begin{equation}
\left.\frac{d\sigma _{AB}^{(H_{A},H_{B})}}{dQ^{2}dy\, d^{n-2}\vec{q}_{T}d\Omega }\right|_{{\mathcal{O}}(\alpha _{S}),\, virtual}=\Phi (\alpha _{S},\varepsilon )\left.\frac{d\sigma _{AB}^{(H_{A},H_{B})}}{dQ^{2}dy\, d^{n-2}\vec{q}_{T}d\Omega }\right|_{LO},\label{virt}\end{equation}
 where\begin{equation}
\Phi (\alpha _{S},\varepsilon )\equiv C_{F}\frac{\alpha _{S}}{2\pi }\left(\frac{4\pi \mu _{F}^{2}}{Q^{2}}\right)^{\varepsilon }\frac{1}{\Gamma (1-\varepsilon )}\left(-\frac{2}{\varepsilon ^{2}}-\frac{3}{\varepsilon }+\pi ^{2}-8+\delta _{v}\right).\label{Phi}\end{equation}
The finite part of $\Phi (\alpha _{S},\eps )$ depends on the prescription
for the $\gamma _{5}$ matrices in $n\neq 4$ dimensions. In the DRED
scheme $\delta _{v}=1$, while in the DREG scheme $\delta _{v}=0$.

The remaining contributions are from the ${\mathcal{O}}(\alpha _{S})$
real emission processes shown in Figs.~\ref{fig:FeynmanDiagrams}e-\ref{fig:FeynmanDiagrams}h.
The real emission cross section can be expressed in terms of the PDFs
$f_{a/A}^{(H_{A})}(\xi _{a},\mu _{F})$, angular functions $A_{\rho }(\theta ,\varphi )$,
and parton-level structure functions ${}^{\rho }T_{ab}(H_{A},H_{B},\shat ,\that ,\uhat )$:
\begin{eqnarray}
 &  & \left.\frac{d\sigma _{AB}^{(H_{A},H_{B})}}{dQ^{2}dy\, dq_{T}^{2}d\Omega }\right|_{{\mathcal{O}}(\alpha _{S}),\, real}=\frac{\alpha _{S}}{2\pi ^{2}s}\sum _{a,b=G,u,\bar{u},d,\bar{d},\dots }\int _{0}^{1}\frac{d\xi _{a}}{\xi _{a}}\int _{0}^{1}\frac{d\xi _{b}}{\xi _{b}}\frac{C_{SU(3)}}{\shat }\delta (\shat +\that +\uhat -Q^{2})\nonumber \\
 &  & \times \, f_{a/A}^{(H_{A})}(\xi _{a},\mu _{F})f_{b/B}^{(H_{B})}(\xi _{b},\mu _{F})\sum _{\rho =-1}^{4}{\mathcal{G}}_{\rho ,ab}^{(H_{A},H_{B})}{}^{\rho }T_{ab}(H_{A},H_{B},\shat ,\that ,\uhat )\, A_{\rho }(\theta ,\varphi ).\label{pert}
\end{eqnarray}
In Eq.~(\ref{pert}), ${\mathcal{G}}_{\rho ,ab}^{(H_{A},H_{B})}$
are the spin-dependent combinations of the constant factors $\sigma _{\ell }^{\pm }$
and $\sigma _{ab}^{\pm }$ defined in Eqs.~(\ref{sigmaab}) and (\ref{sigmal}):
\begin{eqnarray}
{\mathcal{G}}_{\rho ,ab}^{(U,U)}={\mathcal{G}}_{\rho ,ab}^{(P,P)} & \equiv  & \left\{ \begin{array}{cc}
 \sigma _{\ell }^{+}\sigma _{ab}^{+} & \mbox {\quad for\quad }\rho =-1,\, 0\, ,1\, ,2;\\
 \sigma _{\ell }^{-}\sigma _{ab}^{-} & \mbox {\quad for\quad }\rho =3,4;\end{array}
\right.\nonumber \\
{\mathcal{G}}_{\rho ,ab}^{(P,U)} & \equiv  & \left\{ \begin{array}{cc}
 \sigma _{\ell }^{+}\sigma _{ab}^{-} & \mbox {\quad for\quad }\rho =-1,\, 0\, ,1\, ,2;\\
 \sigma _{\ell }^{-}\sigma _{ab}^{+} & \mbox {\quad for\quad }\rho =3,4.\end{array}
\right.
\end{eqnarray}
$C_{SU(3)}$ is a color group factor, which is equal to $C_{F}=(N_{c}^{2}-1)/(2N_{c})=4/3$
in the annihilation subprocess $q\bar{q}\rightarrow GV$ and $T_{R}=1/2$
in Compton scattering subprocess $\stackrel{(-)}{q}G\rightarrow \stackrel{(-)}{q}V$.
The structure functions $^{\rho }T_{ab}(H_{A},H_{B},\shat ,\that ,\uhat )$
in the unpolarized case ($H_A=H_B=U$) can be found in Ref~\cite{CPCsaba}. Our
new result is the calculation of these functions in  
the polarized case ($H_A=P$ and/or $H_B=P$). The explicit expressions
for these functions are given in the appendix.

For any polarization of the proton beams, the structure functions
${}^{-1}T_{ab}(H_{A},H_{B},\shat ,\that ,\uhat )$ and ${}^{3}T_{ab}(H_{A},H_{B},\shat ,\that ,\uhat )$
have a singularity at $q_{T}=0$, which corresponds to the emission
of soft or collinear real partons. The leading-logarithmic part of
the ${\mathcal{O}}(\alpha _{S})$ finite-order cross section (\emph{asymptotic
piece}) at small, but non-zero $q_{T}$ ($0<q_{T}\ll Q$) is \begin{eqnarray}
\left.\frac{d\sigma _{AB}^{(H_{A},H_{B})}}{dQ^{2}dy\, dq_{T}^{2}d\Omega }\right|_{{\mathcal{O}}(\alpha _{S}),q_{T}\rightarrow 0} & \approx  & \frac{1}{s}\frac{\alpha _{S}}{\pi }\frac{1}{2\pi q_{T}^{2}}\sum _{j=u,d,\dots }\, \sum _{\bar{k}=\bar{u},\bar{d},\dots }\Biggl \{\GHAHB \nonumber \\
 & \times  & \biggl [[P_{j/a}^{(H_{A})}\otimes f_{a/A}^{(H_{A})}](x_{A},\mu _{F})f_{\bar{k}/B}^{(H_{B})}(x_{B},\mu _{F})\nonumber \\
 & + & f_{j/A}^{(H_{A})}(x_{A},\mu _{F})[P_{\bar{k}/b}^{(H_{B})}\otimes f_{b/B}^{(H_{B})}](x_{B},\mu _{F})\nonumber \\
 & - & \Bigl (C_{F}\ln \frac{q_{T}^{2}}{Q^{2}}+\frac{3}{2}C_{F}\Bigr )f_{j/A}^{(H_{A})}(x_{A},\mu _{F})f_{\bar{k}/B}^{(H_{B})}(x_{B},\mu _{F})\biggr ]\nonumber \\
 & + & (j\leftrightarrow \bar{k})\Biggr \}.\label{asym}
\end{eqnarray}
 Here the functions $G_{j\bar{k}}^{(H_{A},H_{B})}\left(\theta ,\varphi \right)$
are the same as in Eq.~(\ref{gees}), and $P_{a/b}^{(H_{A})}(\xi )$
are the one-loop splitting functions in Eqs.~(\ref{PUqq})-(\ref{PPqG}).
The convolution of two functions $f_{a/b}(x)$ and $g_{b/c}(x)$ is
defined by\begin{equation}
[f_{a/b}\otimes g_{b/c}](x)\equiv \sum _{b=G,u,\bar{u},d,\bar{d},\dots }\int _{x}^{1}\frac{d\xi }{\xi }f_{a/b}(\frac{x}{\xi })g_{b/c}\left(\xi \right).\end{equation}
This definition includes summation over the repeating parton index
$b$.

In accordance with the Kinoshita-Lee-Nauenberg theorem, the soft singularity,
which appears in the cross section (\ref{asym}) at $q_{T}=0$, is
canceled when this cross section is added to the ${\mathcal{O}}(\alpha _{S})$
virtual contribution (\ref{virt}). Furthermore, the collinear singularities
are absorbed in the parton distributions. The complete small-$q_{T}$
cross section (including both real and virtual corrections) is given
by\begin{eqnarray}
 &  & \left[\left.\frac{d\sigma _{AB}^{(H_{A},H_{B})}}{dQ^{2}dy\, dq_{T}^{2}d\Omega }\right|_{{\mathcal{O}}(\alpha _{S}),q_{T}\rightarrow 0}\right]_{real+virt.}=\left[\left.\frac{d\sigma _{AB}^{(H_{A},H_{B})}}{dQ^{2}dy\, dq_{T}^{2}d\Omega }\right|_{{\mathcal{O}}(\alpha _{S}),q_{T}\rightarrow 0}\right]_{+,\vec{q}_{T}}\nonumber \\
 &  & +\frac{1}{s}\delta ^{2}\left(\vec{q}_{T}\right)\Biggl \{\GHAHB \Biggl [f_{j/A}^{(H_{A})}(x_{A},\mu _{F})f_{\bar{k}/B}^{(H_{B})}(x_{B},\mu _{F})\nonumber \\
 &  & +\frac{\alpha _{S}}{\pi }\Biggl (\Bigl [c_{j/a}^{(H_{A})[1]}\otimes f_{a/A}^{(H_{A})}-\ln \biggl (\frac{\mu _{F}}{Q}\biggr )P_{j/a}^{(H_{A})}\otimes f_{a/A}^{(H_{A})}\Bigr ](x_{A},\mu _{F})f_{\bar{k}/B}^{(H_{B})}(x_{B},\mu _{F})\nonumber \\
 &  & +f_{j/A}^{(H_{A})}(x_{A},\mu _{F})\Bigl [c_{\bar{k}/b}^{(H_{B})[1]}\otimes f_{b/B}^{(H_{B})}-\ln \biggl (\frac{\mu _{F}}{Q}\biggr )P_{\bar{k}/b}^{(H_{B})}\otimes f_{b/B}^{(H_{B})}\Bigr ](x_{B},\mu _{F})\Biggr )\Biggr ]\nonumber \\
 &  & +(j\leftrightarrow \bar{k})\Biggr \}.\label{Oasreg}
\end{eqnarray}
Here the {}``+'' prescription with respect to $\vec{q}_{T}$ (pointed
in the direction of the azimuthal angle $\phi $) is defined as \begin{equation}
\int [f(\vec{q}_{T})]_{+,\vec{q_{T}}}g(\vec{q}_{T})d^{2}\vec{q}_{T}\equiv \int _{0}^{2\pi }d\phi \int _{0}^{+\infty }\left[f(\vec{q}_{T})-f(0)\right]g(\vec{q}_{T})q_{T}dq_{T}.\end{equation}
It acts on the asymptotic piece (\ref{asym}). 

The functions $c_{a/b}^{(H_{A}),[1]}$
are the finite residuals from the cancellation of soft singularities
and factorization of collinear singularities.
In the DREG scheme they are \begin{eqnarray}
\mbox {DREG:\, }c_{i/j}^{(U)[1]}(\xi ) & = & c_{i/j}^{(P)[1]}(\xi )=\delta _{ij}C_{F}\left[\frac{1}{2}(1-\xi )-\left(2-\frac{\pi ^{2}}{4}\right)\delta (1-\xi )\right];\\
\mbox {\, \, \, \, \, }c_{i/G}^{(U)[1]}(\xi ) & = & \frac{1}{2}\xi (1-\xi );\\
\mbox {\, \, \, \, \, }c_{i/G}^{(P)[1]}(\xi ) & = & \frac{1}{2}(1-\xi );\, i,j=u,\bar{u},d,\bar{d},\dots 
\end{eqnarray}
In the DRED scheme, these functions are
\begin{eqnarray}
\mbox {DRED:\, }c_{i/j}^{(U)[1]}(\xi ) & = & c_{i/j}^{(P)[1]}(\xi )=-\delta _{ij}C_{F}\left[\frac{7}{4}-\frac{\pi ^{2}}{4}\right]\delta (1-\xi );\\
\mbox {\, \, \, \, \, }c_{i/G}^{(U)[1]}(\xi ) & = & c_{i/G}^{(P)[1]}(\xi )=0.
\end{eqnarray}
It can be seen from Eqs.\,(\ref{asym}) and (\ref{Oasreg}) that the dependence
on the the lepton decay angles $\theta$ and $\phi$ in the small-$q_T$ limit
is   contained entirely in the overall factor $G_{j\bar
  k}^{(H_A,H_B)}(\theta,\phi)$, which is exactly the same as in
  the leading-order cross section (\ref{LO}).   
Such factorization of the dependence on $\theta$ and $\phi$ 
from the hadronic dynamics reflects preservation of the chirality structure
of the electroweak vertex as $q_T \rightarrow 0$, i.e.,
the Born-level angular distributions of the leptons are not affected by
the soft or collinear parton
radiation. This feature, in its turn, automatically follows from the 
use of the DRED scheme in our calculation. 
 
\subsection{Resummation of large logarithms}

The ${\mathcal{O}}(\alpha _{S})$ cross section (\ref{Oasreg}) is
given in terms of generalized functions and can be used to predict
the integrated rate. For instance, the integrated rate over the region
$0\leq q_{T}^{2}\leq P_{T}^{2}$ (where $P_{T}$ is small enough for
the approximation (\ref{asym}) to be valid) is \begin{eqnarray}
 &  & \int _{0}^{P_{T}^{2}}dq_{T}^{2}\frac{d\sigma _{AB}^{(H_{A},H_{B})}}{dQ^{2}dy\, dq_{T}^{2}d\Omega }=\frac{1}{s}\sum _{j,\bar{k}}\Biggl \{\GHAHB \nonumber \\
 &  & \times \Biggl [\left(1-\frac{\alpha _{S}}{\pi }\left(\frac{1}{2}C_{F}\ln ^{2}\frac{Q^{2}}{P_{T}^{2}}-\frac{3}{2}C_{F}\ln \frac{Q^{2}}{P_{T}^{2}}\right)\right)f_{j/A}^{(H_{A})}(x_{A},\mu _{F})f_{\bar{k}/B}^{(H_{B})}(x_{B},\mu _{F})\nonumber \\
 &  & -\frac{\alpha _{S}}{2\pi }\ln \left(\frac{\mu _{F}^{2}}{P_{T}^{2}}\right)\Biggl (\biggl [P_{j/a}^{(H_{A})}\otimes f_{a/A}^{(H_{A})}\biggr ](x_{A},\mu _{F})f_{\bar{k}/B}^{(H_{B})}(x_{B},\mu _{F})\nonumber \\
 &  & +f_{j/A}^{(H_{A})}(x_{A},\mu _{F})\biggl [P_{\bar{k}/b}^{(H_{B})}\otimes f_{b/B}^{(H_{B})}\biggr ](x_{B},\mu _{F})\Biggl )\nonumber \\
 &  & +\frac{\alpha _{S}}{\pi }\Biggl (\biggl [c_{ja}^{(H_{A})[1]}\otimes f_{a/A}^{(H_{A})}\biggr ](x_{A},\mu _{F})f_{\bar{k}/B}^{(H_{B})}(x_{B},\mu _{F})+\nonumber \\
 &  & +f_{j/A}^{(H_{A})}(x_{A},\mu _{F})\biggl [c_{\bar{k}b}^{(H_{B})[1]}\otimes f_{b/B}^{(H_{B})}\biggr ](x_{B},\mu _{F})\Biggr )\Biggr ]+(j\leftrightarrow \bar{k})\Biggr \}.\label{DelSig}
\end{eqnarray}
However, Eq.~(\ref{Oasreg}) cannot be used yet on its own to describe
the distribution over $q_{T}$. To obtain a meaningful distribution
in the small-$q_{T}$ region, we proceed further and identify Eq.~(\ref{Oasreg})
as an ${\mathcal{O}}(\alpha _{S})$ part in the perturbative expansion
of the small-$q_{T}$ resummed cross section \cite{CSS}, given by
\begin{eqnarray}
\frac{d\sigma _{AB}^{(H_{A},H_{B})}}{dQ^{2}dydq_{T}^{2}d\Omega } & = & \frac{1}{s}\int \frac{d^{2}\vec{b}}{(2\pi )^{2}}e^{i\vec{q}_{T}\cdot \vec{b}}\, \widetilde{W}_{AB}^{(H_{A},H_{B})}(Q,b,y,\Omega )+Y_{AB}^{(H_{A},H_{B})}(Q,q_{T},y,\Omega ).\label{WY}
\end{eqnarray}

In Eq.~(\ref{WY}), the first term on the right-hand side dominates
when $q_{T}\rightarrow 0$. It is expressed as a Fourier-Bessel transform
of a form factor $\widetilde{W}_{AB}^{(H_{A},H_{B})}(Q,b,y,\Omega )$
in the space of the impact parameter $b$ (conjugate to $q_{T}$):\begin{eqnarray}
 &  & \widetilde{W}_{AB}^{(H_{A},H_{B})}(Q,b,y,\Omega )=\sum _{j=u,d,\dots }\, \sum _{\bar{k}=\bar{u},\bar{d},\dots }\Biggl [\GHAHB \exp {\left(-\S ^{(H_{A},H_{B})}(Q,b,y,C_{1},C_{2})\right)}\nonumber \\
 &  & \times \biggl [\C _{j/a}^{(H_{A})}\otimes f_{a/A}^{(H_{A})}\biggr ]\bigl (x_{A},b_{*},\frac{C_{1}}{C_{2}},\mu =\frac{C_{3}}{b_{*}}\bigr )\biggl [\C _{\bar{k}/b}^{(H_{B})}\otimes f_{b/B}^{(H_{B})}\biggr ]\bigl (x_{B},b_{*},\frac{C_{1}}{C_{2}},\mu =\frac{C_{3}}{b_{*}}\bigr )\nonumber \\
 &  & +(j\leftrightarrow \bar{k})\Biggr ].\label{W}
\end{eqnarray}
Here $C_{1}$ and $C_{2}$ are constants of order unity that determine
the momentum scales separating the exponential soft factor $e^{-\S }$
from the convolutions $\biggl [{\mathcal{C}}_{a/b}^{(H)}\otimes f_{b/A}^{(H)}\biggr ](x,b,C_{1}/C_{2},\mu ).$
The constant $C_{3}$ specifies the factorization scale $\mu ,$ which
separates the function ${\mathcal{C}}_{a/b}^{(H)}(x,b,C_{1}/C_{2},\mu =\nolinebreak [4]C_{3}/b)$
from the PDF $f_{a/A}^{(H)}(x,\mu =C_{3}/b)$ in the convolution $\biggl [{\mathcal{C}}_{a/b}^{(H)}\otimes f_{b/A}^{(H)}\biggr ](x,b,C_{1}/C_{2},\mu =C_{3}/b).$
For definiteness, the form factor $\widetilde{W}_{AB}^{(H_{A},H_{B})}(Q,b,y,\Omega )$
in Eq.~(\ref{W}) is written in its common (and, perhaps, the simplest)
form proposed in Ref.~\cite{CSS}. Namely, the perturbative parts
of $\widetilde{W}_{AB}^{(H_{A},H_{B})}(Q,b,y,\Omega )$ are evaluated
at a variable\begin{equation}
b_{*}=\frac{b}{\sqrt{1+b^{2}/b_{max}^{2}}},\label{bstar}\end{equation}
where the parameter $b_{max}$ is of order $1\mbox {\, GeV}^{-1}$.
The variable $b_{*}$ is introduced to separate short- and long-distance
dynamics in $\widetilde{W}_{AB}^{(H_{A},H_{B})}(Q,b,y,\Omega )$.
Furthermore, $\widetilde{W}_{AB}^{(H_{A},H_{B})}(Q,b,y,\Omega )$
is normalized to the Born-level hard contribution (included in the
normalization factor $G_{j\bar{k}}^{(H_{A},H_{B})}(\theta ,\varphi )$),
while higher-order corrections to the hard part are absorbed in the
functions $[{\mathcal{C}}\otimes f]$ and $e^{-\S }$. Recently discussed
alternatives to this representation can be found in Refs.~\cite{deFlorianPRL2000,CataniDeFlorianGrazzini2000,Qiu:2000ga,Qiu:2000hf}.

In this representation, the soft (Sudakov) factor $\S ^{(H_{A},H_{B})}(b,Q,y)$
consists of the perturbative part $\S _{P}(Q,b,C_{1},C_{2})$ and
the nonperturbative part $\S _{NP}^{(H_{A},H_{B})}(Q,b,y)$:\begin{equation}
\S ^{(H_{A},H_{B})}(Q,b,y)=\S _{P}(Q,b_{*},C_{1},C_{2})+\S _{NP}^{(H_{A},H_{B})}(Q,b,y),\label{Sudakov}\end{equation}
where\begin{eqnarray}
\S _{P}(Q,b,C_{1},C_{2}) & = & \int _{C_{1}^{2}/b^{2}}^{C_{2}^{2}Q^{2}}\frac{d\overline{\mu }^{2}}{\overline{\mu }^{2}}\Biggl (\A \biggl (\alpha _{s}(\overline{\mu }),C_{1}\biggr )\ln \frac{C_{2}^{2}Q^{2}}{\overline{\mu }^{2}}+\B \biggl (\alpha _{s}(\overline{\mu }),C_{1},C_{2}\biggr )\Biggr ).\label{SP}
\end{eqnarray}
The functions $\A $ and $\B $ in $\S _{P}$ can be calculated in
perturbation theory, as\begin{eqnarray}
\A \biggl (\alpha _{S}(\bar{\mu }),C_{1}\biggr ) & = & \sum _{m=1}^{\infty }\left(\frac{\alpha _{s}(\overline{\mu })}{\pi }\right)^{m}\A ^{[m]}(C_{1}),\\
\B \biggl (\alpha _{s}(\overline{\mu }),C_{1},C_{2}\biggr ) & = & \sum _{m=1}^{\infty }\left(\frac{\alpha _{s}(\overline{\mu })}{\pi }\right)^{m}\B ^{[m]}(C_{1},C_{2}).
\end{eqnarray}
The coefficients in these expansions are known up to order ${\mathcal{O}}(\alpha _{S}^{2})$.
Explicitly, they are\begin{eqnarray}
\A ^{[1]} & = & C_{F},\label{A1}\\
\A ^{[2]} & = & C_{F}\Biggl [\left(\frac{67}{36}-\frac{\pi ^{2}}{12}\right)N_{c}-\frac{5}{18}N_{F}-\beta _{0}\ln \left(\frac{b_{0}}{C_{1}}\right)\Biggr ],\label{A2}\\
\B ^{[1]} & = & 2C_{F}\ln \left(\frac{e^{-3/4}C_{1}}{b_{0}C_{2}}\right),\label{B1}\\
\B ^{[2]} & = & C_{F}\Biggl \{C_{F}\left(\frac{\pi ^{2}}{4}-\frac{3}{16}-3\zeta (3)\right)+N_{c}\left(\frac{11}{36}\pi ^{2}-\frac{193}{48}+\frac{3}{2}\zeta (3)\right)\nonumber \\
 & + & N_{F}\left(-\frac{1}{18}\pi ^{2}+\frac{17}{24}\right)-\left[N_{c}\left(\frac{67}{18}-\frac{\pi ^{2}}{6}\right)-\frac{5}{9}N_{F}\right]\ln \left(\frac{C_{2}b_{0}}{C_{1}}\right)\nonumber \\
 & + & \beta _{0}\left[\ln ^{2}\left(\frac{b_{0}}{C_{1}}\right)-\ln ^{2}C_{2}-\frac{3}{2}\ln C_{2}\right]\Biggr \}.\label{B2}
\end{eqnarray}
In these equations, $N_{F}$ is the number of active quark flavors,
$\beta _{0}=\left(11N_{c}-2N_{F}\right)/6$, $\zeta (3)\approx 1.202$
is the Riemann zeta function, and $b_{0}=2e^{-\gamma _{E}}$ is a
function of the Euler constant $\gamma _{E}\approx 0.577$. In the
present work, we have re-derived the coefficients $\A ^{[1]},\, \B ^{[1]}$,
and they agree with the earlier published results \cite{Weber1,Weber2}.
The coefficient $\A ^{[2]}$ for the spin-dependent collisions was
discussed earlier in \cite{Weber1}; since it comes from the leading-logarithmic
contributions generated by the soft gluons, it is expected to be independent
of spin.%
\footnote{The spin independence of soft contributions is evident from the Feynman
rules for the soft eikonal approximation; see, for instance, Ref.~\cite{CollinsSoper81}.%
} 

We have also determined the coefficient $\B ^{[2]}$ in Eq.~(\ref{B2})
by employing the method of Refs.$\, $\cite{deFlorianPRL2000,CataniDeFlorianGrazzini2000,deFlorian:2001zd}
and order-by-order independence of the full form factor $\widetilde{W}_{AB}^{(H_{A},H_{B})}$
from the choice of the scales $C_{1}/b$ and $C_{2}Q.$ By utilizing
the independence on $C_{1}$ and $C_{2}$, we find\begin{equation}
\B ^{[2]}=\B _{c}^{[2]}+\beta _{0}\left(\A _{c}^{[1]}\ln ^{2}\left(\frac{b_{0}}{C_{1}}\right)+\B _{c}^{[1]}\ln C_{2}-\A _{c}^{[1]}\ln ^{2}C_{2}\right)-\A _{c}^{[2]}\ln \left(\frac{b_{0}^{2}C_{2}^{2}}{C_{1}^{2}}\right),\end{equation}
where the subscript $"c"$ denotes the coefficients in Eqs.~(\ref{A1})-(\ref{B2})
evaluated for the {}``canonical'' variables $C_{1}=b_{0}$ and $C_{2}=1$:\begin{eqnarray}
\A _{c}^{[1]} & = & C_{F},\, \, \B _{c}^{[1]}=-\frac{3}{2}C_{F},\\
\A _{c}^{[2]} & = & C_{F}N_{c}\left(\frac{67}{36}-\frac{\pi ^{2}}{12}\right)N_{c}-\frac{5}{18}C_{F}N_{F},
\end{eqnarray}
and\begin{equation}
\B _{c}^{[2]}=C_{F}^{2}\left(\frac{\pi ^{2}}{4}-\frac{3}{16}-3\zeta (3)\right)+N_{c}C_{F}\left(\frac{11}{36}\pi ^{2}-\frac{193}{48}+\frac{3}{2}\zeta (3)\right)+C_{F}N_{F}\left(-\frac{1}{18}\pi ^{2}+\frac{17}{24}\right).\end{equation}
To derive the last expression, we use the generalization of the relationship
from \cite{deFlorianPRL2000,CataniDeFlorianGrazzini2000}, i.e., \begin{equation}
\B _{c}^{[2]}=-\frac{\gamma _{A}^{[2]}+\gamma _{B}^{[2]}}{4}+\beta _{0}\left(\frac{{\mathcal{V}}_{qq}}{4}+\frac{C_{F}\pi ^{2}}{12}\right).\end{equation}
This equation relates the canonical value of the coefficient $\B ^{[2]}$
to the finite contribution ${\mathcal{V}}_{qq}/4\equiv C_{F}(\pi ^{2}-8)/4$
from the ${\mathcal{O}}(\alpha _{S})$ virtual correction $\Phi (\alpha _{S},\eps )$
and coefficients $\gamma _{A}^{[2]}$, $\gamma _{B}^{[2]}$ of the
$\delta (1-x)$ terms in the two-loop $qq$ splitting functions on
the sides of the hadrons $A$ and $B$, respectively. Due to the helicity
conservation in the quark-gluon vertex for massless quarks, the $\delta (1-x)$
terms are the same in the unpolarized and polarized case, so that

\begin{equation}
\gamma _{A}^{[2]}=\gamma _{B}^{[2]}=C_{F}^{2}\left(\frac{3}{8}-\frac{\pi ^{2}}{2}6\zeta (3)\right)+C_{F}N_{c}\left(\frac{17}{24}+\frac{11}{18}\pi ^{2}-3\zeta (3)\right)-\frac{1}{2}C_{F}N_{F}\left(\frac{1}{6}+\frac{2\pi ^{2}}{9}\right).\end{equation}
Hence, similarly to $\A ^{[2]}$, the coefficient $\B ^{[2]}$ in
Eq.~(\ref{SP}) is also independent of the spin of the incoming quarks.

Alternatively, the spin independence of the coefficient $\B ^{[2]}$
can be shown in the anticommuting $\gamma _{5}$ or the DRED schemes
by analyzing the spin structure of the ${\mathcal{O}}(\alpha _{S}^{2})$
cut diagrams with the radiation of one or two gluons. The coefficient
$\B ^{[2]}$ is generated by the diagrams that reduce to the Born
level diagram in the limit when the unobserved radiated partons are
soft. All such cut diagrams have only one closed fermion line and
can be reduced to the corresponding unpolarized diagram (up to an
overall sign) by commutation of the $\gamma _{5}$ matrices from the
fermion spin projectors along the fermion line. Therefore, they lead
to the same expression for $\B ^{[2]}$ as in the unpolarized case.%
\footnote{We thank W. Vogelsang for pointing out the alternative proof of the
spin independence of the coefficient~${\mathcal{B}}^{[2]}$.%
} To summarize, our study confirms the spin independence of all Sudakov
coefficients in Eqs.~(\ref{A1})-(\ref{B2}), which agree with their
values known from the unpolarized calculations \cite{CSS,KodairaA2,DaviesStirling84,DWS}.

Similarly to the perturbative Sudakov factor, the functions $\C _{a/b}^{(H)}(\xi ,b,\mu )$
can be calculated order by order in PQCD:\begin{eqnarray}
\C _{a/b}^{(H)}(\xi ,b,\mu ) & = & \sum _{m=0}^{\infty }\C _{a/b}^{(H)[m]}(\xi ,b\mu )\left(\frac{\alpha _{s}(\mu )}{\pi }\right)^{m},\quad H=U,P.
\end{eqnarray}
The coefficients $\C ^{(H)[m]}$ are known up to order ${\mathcal{O}}(\alpha _{S})$.
The lowest-order coefficients are trivial:\begin{eqnarray}
\C _{i/j}^{(U)[0]}(\xi ,b\mu ) & = & \C _{i/j}^{(P)[0]}(\xi ,b\mu )=\delta _{ij}\delta (1-\xi );\\
\C _{i/G}^{(H)[0]} & = & \C _{G/i}^{(H)[0]}=\C _{G/G}^{(H)[1]}=0;\, i,j=u,\bar{u},d,\bar{d},\dots \, .
\end{eqnarray}
All non-zero ${\mathcal{O}}(\alpha _{S})$ coefficients in the DRED
and DREG schemes are given by Eqs.~(\ref{C1HijDRED}),(\ref{C1HiGDRED}),
and (\ref{C1HijDREG})-(\ref{C1PiGDREG}). As discussed in Section~\ref{sec:DREGandDRED},
the coefficients $\C _{j/a}^{(H)[1]}(\xi ,b\mu )$ depend on the prescription
for continuation of the $\gamma _{5}$ matrix to $n\neq 4$ dimensions,
and this dependence is entirely determined by the form of the $n$-dimensional
splitting functions $P_{a/b}^{(H),\eps }(\xi )$ in the DRED and DREG
schemes given in Eqs.~(\ref{PHepsDRED}) and (\ref{PHepsDREG1})-(\ref{PHepsDREG4}).
The above equations explicitly demonstrate the factorization of the
collinear contributions associated with the hadron $A$ from the collinear
contributions associated with the hadron $B$. That is, the $\C $-function
associated with the hadron $A$ depends only on the polarization of
$A$ and does not depend on the polarization of the hadron $B$; and
vice versa. Such factorization completely agrees with the general
structure of the resummed cross section, in which the part associated
with the collinear radiation along the beam $A$ is independent from
the collinear radiation along the beam $B$.

Our results for the functions $\C _{i/G}^{(U)}(\xi ,b,\mu )$ and
$\C _{i/G}^{(P)}(\xi ,b,\mu )$ differ from the corresponding expressions
in the earlier publications \cite{Weber1} and \cite{Weber2}. Firstly,
in accordance with the factorization of collinear contributions, our
expression for the unpolarized function $\C _{i/G}^{(U)}(\xi ,b,\mu )$
is the same in the unpolarized and single-spin processes. On the other
hand, Ref.~\cite{Weber2} has found $\C _{i/G}^{(U)[1]}(\xi ,b,\mu )$
in single-spin $W$ boson production to be\begin{equation}
\left.\C _{i/G}^{(U)[1]}(\xi ,b,\mu )\right|_{\mbox {\small Ref.\, {\cite {Weber2}}}}=\frac{1}{4}-P_{q/G}^{(U)}(\xi )\ln \Bigl (\frac{\mu b}{b_{0}}\Bigr ),\label{C1UigWeber}\end{equation}
so that the above factorization does not hold.%
\footnote{Note that Refs.~\cite{Weber1,Weber2} expand in the series of $(\alpha _{S}/2\pi )$,
so that the ${O}(\alpha _{S})$ coefficients in those papers are two
times larger than the $O(\alpha _{S}/\pi )$ coefficients presented
here.%
} This difference may be caused by the choice of the normalization
of the $q^{(P)}G^{(U)}$ cross section in Ref.~\cite{Weber2}. Namely,
the same result as in Eq.~(\ref{C1UigWeber}) would be obtained if
the $q^{(P)}G^{(U)}$ cross section in $n=4-2\eps $ dimensions were
averaged over two polarizations of the initial gluon and not over
$2(1-\eps )$ polarizations, as it is required for an unpolarized
initial-state gluon in the $\overline{MS}$ scheme. 

Secondly, Refs.~\cite{Weber1,Weber2} have used a different factorization
scheme for the polarized gluon-initiated functions $\C _{i/G}^{(P)}(\xi ,b,\mu )$.
In that scheme, $\C _{i/G}^{(P)[1]}(\xi ,b\mu )$ is identically zero
for the {}``canonical'' scale $\mu =b_{0}/b$: \begin{equation}
\left.\C _{i/G}^{(P)[1]}(\xi ,b\mu )\right|_{\mbox {\small Refs.\, {\cite {Weber1,Weber2}}}}=-P_{q/G}^{(P)}(\xi )\ln \Bigl (\frac{\mu b}{b_{0}}\Bigr ),\end{equation}
so that $\C _{i/G}^{(P)[1]}(\xi ,b\mu )=0$ if $\mu =b_{0}/b$. To
obtain such $\C -$function, the ${\mathcal{O}}(\alpha _{S})$ quark
distribution in the gluon has to be\begin{equation}
\left.f_{q/G}^{(P)}(\xi ,\mu )\right|_{\mbox {\small Refs.\, {\cite {Weber1,Weber2}}}}=\frac{\alpha _{S}}{2\pi }\left(-\frac{1}{\eps }P_{q/G}^{(P)}(\xi )+1-\xi \right)+{\mathcal{O}}\left(\alpha _{S}^{2}\right),\end{equation}
which is different from $f_{q/G}^{(P)}(\xi ,\mu )=-(\alpha _{S}/2\pi \eps )P_{q/G}^{(P)}(\xi )+{\mathcal{O}}(\alpha _{S}^{2})$
in the $\overline{MS}$ scheme. While this definition is not contradictory,
it is not the same as the convention of the $\overline{MS}$ scheme
\cite{Mertig:1996ny,Vogelsang:1996vh,VogelsangTwoLoop}. Consequently,
the functions $\C _{i/G}^{(U)[1]}(\xi ,b\mu )$ and $\C _{i/G}^{(P)[1]}(\xi ,b\mu )$
found in Refs.~\cite{Weber1,Weber2} cannot be combined with the
$\overline{MS}$ parton distributions. Instead, a calculation in the
$\overline{MS}$ scheme must use the functions $\C _{a/b}^{(H)[1]}(\xi ,b\mu )$
presented in Eqs.~(\ref{C1HijDREG})-(\ref{C1PiGDREG}).

The remaining part of the form-factor $\widetilde{W}_{AB}^{(H_{A},H_{B})}(Q,b,y,\Omega )$
to be discussed is the nonperturbative Sudakov function $S_{NP}(b,Q)$
(see Eq.~(\ref{Sudakov})), which determines the behavior of this
form factor at $b\gtrsim 1\mbox {\, GeV}^{-1}$. While at the present
stage it is not possible to calculate this nonperturbative function
from the first principles, its phenomenological parametrization can
be found from the global analysis of VBP data. At the time of writing
of this paper, the latest parametrization of $S_{NP}(b,Q)$ in the
$b_{*}$ approach in the unpolarized case is available in Ref.~\cite{Landry:2002ix}.
Finally, we comment on the finite term, $Y(Q,q_{T},y,\Omega )$, in
Eq.~(\ref{WY}). It is given by the difference between the ${\mathcal{O}}(\alpha _{S})$
finite-order cross section and its singular part (asymptotic piece)
that is already included in the $\wt{W}$ term:\begin{equation}
Y_{AB}^{(H_{A},H_{B})}(Q,q_{T},y,\Omega )\equiv \left.\frac{d\sigma _{AB}^{(H_{A},H_{B})}}{dQ^{2}dydq_{T}^{2}d\Omega }\right|_{{\mathcal{O}}\mathcal{(}\alpha _{S}\mathcal{)}}-\left.\frac{d\sigma _{AB}^{(H_{A},H_{B})}}{dQ^{2}dy\, dq_{T}^{2}d\Omega }\right|_{{\mathcal{O}}(\alpha _{S}),q_{T}\rightarrow 0}.\end{equation}
The ${\mathcal{O}}(\alpha _{S})$ perturbative cross section for non-zero
$q_{T}$ is given in Eq.~(\ref{pert}), while the asymptotic piece
is given in Eq.~(\ref{asym}). This term is regular and, in fact,
vanishes as $q_{T}\rightarrow 0$, so that the distribution (\ref{WY})
in the small-$q_{T}$ region is approximated well by the $\widetilde{W}$
term. At $q_{T}\gtrsim Q$, the cross section (\ref{WY}) reduces
to the finite-order cross section (\ref{pert}), up to higher-order
corrections.

\section{Summary\label{sec:Summary}}
In this paper, we presented fully differential distributions for production
and decay of electroweak bosons ($\gamma^*,\ W^\pm,\ Z^0$) 
in the collisions of proton beams of arbitrary
longitudinal polarizations. 
One of our new results is the complete set of spin-dependent 
structure functions of order ${\mathcal{O}}(\alpha _{S})$  for 
the angular distributions of leptons from the vector boson decay.
This finite-order
cross section is combined with the all-order sum of leading logarithms
in the small transverse momentum region using the Collins-Soper-Sterman
(impact parameter) resummation method \cite{CSS}. The perturbative
coefficients are presented at the one-loop level, with the perturbative
Sudakov factor presented at the two-loop level. For the first time,
we have explicitly demonstrated the universality of QCD factorization in
the perturbative part of the spin-dependent resummed cross section. Namely, 
the soft factor is independent of spin, while the form of 
the {}``$b$-dependent PDFs'' $[{\mathcal{C}}_{a/b}\otimes f_{b/A}](x,b,\mu )$ 
is entirely determined by the type and polarization of the 
corresponding initial-state
hadron, and it is independent from the type and polarization of the
other initial-state hadron. We have also demonstrated the spin independence
of the ${\mathcal{O}}(\alpha _{S}^{2})$ coefficient ${\cal B}^{(2)}$
based on the method of 
Refs.~\cite{deFlorianPRL2000,CataniDeFlorianGrazzini2000},
and, alternatively, on the analysis of the spin structure of 
${\mathcal{O}}(\alpha _{S}^{2})$
Feynman diagrams. The spin-dependent cross sections were presented
in the dimensional regularization \cite{'tHooft:1972fi,Breitenlohner:1977hr}
and dimensional reduction \cite{Siegel:1979wq} schemes. 

The effects of soft parton radiation and vector boson decay considered
here must be understood well in order to correctly describe 
production of Drell-Yan pairs and massive electroweak bosons 
at the Relativistic Heavy Ion Collider. 
In the accompanying paper \cite{PolWPhenomenology}, we 
use these results to study measurements of sea quark parton distributions 
in weak boson production with longitudinally polarized proton beams.

\section*{Acknowledgments}

Authors would like to thank C. Balazs, D. Boer, C. Glosser, X.~Ji,
S.~Kretzer, N. Saito, C.~Schmidt, T.~Tait, W.~Vogelsang, and the
members of CTEQ Collaboration for useful discussions. Authors also
thank the organizers of RHIC Spin workshops, where the preliminary
results of this work were presented. The work of P.M.N. has been supported
by the U.S. Department of Energy and Lightner-Sams Foundation. The
research of C.P.Y. has been supported by the National Science Foundation
under grant PHY-0100677.

\appendix

\section*{\label{AppendixStructureFns}Appendix: The structure functions for
the finite-order cross section}

This appendix collects explicit expressions for the structure functions
${}^{\rho }T_{ab}(H_{A},H_{B},\shat ,\that ,\uhat )$ that enter the
${\mathcal{O}}(\alpha _{S})$ real emission cross section in Eq.~(\ref{pert}).
Eqs.~(\ref{TqqUU}-\ref{TqGPP}) show the independent structure functions
${}^{\rho }\Tqq ,\, {}^{\rho }T_{Gq},\, ,$ and, in the case of the
single-spin cross section, ${}^{\rho }T_{qG}$. The end of the appendix
includes the rules to derive the rest of nonzero structure functions.
As a convenient notation, the mathematical expressions will contain
the functions $T_{\pm }(a,b)$, defined as\begin{equation}
T_{\pm }(a,b)\equiv \Biggl (\left(Q^{2}-a\right)^{2}\pm \left(Q^{2}-b\right)^{2}\Biggr ).\end{equation}

The independent structure functions in the unpolarized cross section
are as follows \cite{CPCsaba}. In the $q^{(U)}(p_{a})\bar{q}^{(U)}(p_{b})$
subprocess,\begin{eqnarray}
^{-1}T_{q\bar{q}} & = & \frac{T_{+}(\uhat ,\that )}{q_{T}^{2}},\label{-1TqqUU}\nonumber \\
^{0}\Tqq  & = & ^{2}\Tqq =\frac{T_{+}(\uhat ,\that )}{M_{T}^{2}},\nonumber \\
^{1}\Tqq  & = & \frac{Q}{q_{T}}\frac{T_{-}(\uhat ,\that )}{M_{T}^{2}},\nonumber \\
^{3}\Tqq  & = & \frac{Q}{M_{T}}\frac{T_{+}(\uhat ,\that )}{q_{T}^{2}},\nonumber \\
^{4}\Tqq  & = & 2\frac{T_{-}(\uhat ,\that )}{M_{T}q_{T}}.\label{TqqUU}
\end{eqnarray}
In the $G^{(U)}(p_{a})q^{(U)}(p_{b})$ subprocess,\begin{eqnarray}
^{-1}T_{Gq} & = & -\frac{T_{+}(\shat ,\uhat )}{q_{T}^{2}}\frac{\that }{\shat },\nonumber \\
^{0}\TGq  & = & ^{2}\TGq =-\frac{T_{+}(-\shat ,\uhat )}{M_{T}^{2}}\frac{\that }{\shat },\nonumber \\
^{1}\TGq  & = & -\frac{Q}{q_{T}}\frac{T_{-}(\uhat ,\that )+\left(Q^{2}-\uhat \right)^{2}}{M_{T}^{2}}\frac{\that }{\shat },\nonumber \\
^{3}\TGq  & = & \frac{Q}{M_{T}}\frac{T_{+}(\uhat ,\shat )-2\uhat (Q^{2}-\shat )}{q_{T}^{2}}\frac{\that }{\shat },\nonumber \\
^{4}\TGq  & = & 2\frac{T_{+}(\uhat ,\shat )+2\left(Q^{2}-\shat \right)\shat }{M_{T}q_{T}}\frac{\that }{\shat }.\label{TqGUU}
\end{eqnarray}

Next, the independent structure functions in the single-spin cross
section are as follows. In the $q^{(P)}(p_{a})\bar{q}^{(U)}(p_{b})$
subprocess,\begin{eqnarray}
^{-1}\Tqq  & = & -\frac{T_{+}(\uhat ,\that )}{q_{T}^{2}},\nonumber \\
^{0}\Tqq  & = & {}^{2}\Tqq =-\frac{T_{+}(\uhat ,\that )}{M_{T}^{2}},\nonumber \\
^{1}\Tqq  & = & -\frac{Q}{q_{T}}\frac{T_{-}(\uhat ,\that )}{M_{T}^{2}},\nonumber \\
^{3}\Tqq  & = & -\frac{Q}{M_{T}}\frac{T_{+}(\uhat ,\that )}{q_{T}^{2}},\nonumber \\
^{4}\Tqq  & = & -2\frac{T_{-}(\uhat ,\that )}{M_{T}q_{T}}.\label{TqqPU}
\end{eqnarray}
In the $q^{(P)}(p_{a})G^{(U)}(p_{b})$ subprocess,\begin{eqnarray}
^{-1}\TqG  & = & \frac{T_{+}(\that ,\shat )}{q_{T}^{2}}\frac{\uhat }{\shat },\nonumber \\
^{0}\TqG  & = & {}^{2}\TqG =\frac{T_{+}(\that ,-\shat )}{M_{T}^{2}}\frac{\uhat }{\shat },\nonumber \\
^{1}\TqG  & = & -\frac{Q}{q_{T}}\frac{T_{-}(\that ,\uhat )+(Q^{2}-\that )^{2}}{M_{T}^{2}}\frac{\uhat }{\shat },\nonumber \\
^{3}\TqG  & = & \frac{Q}{M_{T}}\frac{T_{+}(\that ,\shat )-2\left(Q^{2}-\shat \right)\that }{q_{T}^{2}}\frac{\uhat }{\shat },\nonumber \\
^{4}\TqG  & = & -2\frac{T_{+}(\that ,\shat )+2(Q^{2}-\shat )\shat }{M_{T}q_{T}}\frac{\uhat }{\shat }.\label{TqGPU}
\end{eqnarray}
In the $G^{(P)}(p_{a})q^{(U)}(p_{b})$subprocess,\begin{eqnarray}
^{-1}\TGq  & = & -\frac{T_{-}(\uhat ,\shat )}{q_{T}^{2}}\frac{\that }{\shat },\nonumber \\
^{0}\TGq  & = & {}^{2}\TGq =-\frac{T_{-}(\uhat ,-\shat )}{M_{T}^{2}}\frac{\that }{\shat },\nonumber \\
^{1}\TGq  & = & \frac{Q}{q_{T}}\frac{T_{-}(\uhat ,\that )-(Q^{2}-\uhat )^{2}}{M_{T}^{2}}\frac{\that }{\shat },\nonumber \\
^{3}\TGq  & = & \frac{Q}{M_{T}}\frac{T_{-}(\uhat ,\shat )+2\uhat (Q^{2}-\shat )}{q_{T}^{2}}\frac{\that }{\shat },\nonumber \\
^{4}\TGq  & = & 2\frac{T_{-}(\uhat ,\shat )-2\shat (Q^{2}-\shat )}{M_{T}q_{T}}\frac{\that }{\shat }.\label{TGqPU}
\end{eqnarray}

Finally, the independent structure functions in the double-spin cross
section are as follows. In the $q^{(P)}(p_{a})\bar{q}^{(P)}(p_{b})$
subprocess,\begin{eqnarray}
{}^{-1}\Tqq  & = & -\frac{T_{+}(\uhat ,\that )}{q_{T}^{2}},\nonumber \\
^{0}\Tqq  & = & {}^{2}\Tqq =-\frac{T_{+}(\uhat ,\that )}{M_{T}^{2}},\nonumber \\
^{1}\Tqq  & = & -\frac{Q}{q_{T}}\frac{T_{-}(\uhat ,\that )}{M_{T}^{2}},\nonumber \\
^{3}\Tqq  & = & -\frac{Q}{M_{T}}\frac{T_{+}(\uhat ,\that )}{q_{T}^{2}},\nonumber \\
^{4}\Tqq  & = & -2\frac{T_{-}(\uhat ,\that )}{M_{T}q_{T}}.\label{TqqPP}
\end{eqnarray}
In the $G^{(P)}(p_{a})q^{(P)}(p_{b})$ subprocess,\begin{eqnarray}
^{-1}\TGq  & =- & \frac{T_{-}(\shat ,\uhat )}{q_{T}^{2}}\frac{\that }{\shat },\nonumber \\
^{0}\TqG  & = & {}^{2}\TGq =-\frac{T_{-}(-\shat ,\uhat )}{M_{T}^{2}}\frac{\that }{\shat },\nonumber \\
^{1}\TGq  & = & -\frac{Q}{q_{T}}\frac{T_{-}(\uhat ,\that )-(Q^{2}-\uhat )^{2}}{M_{T}^{2}}\frac{\that }{\shat },\nonumber \\
^{3}\TGq  & = & -\frac{Q}{M_{T}}\frac{T_{-}(\uhat ,\shat )+2\uhat (Q^{2}-\shat )}{q_{T}^{2}}\frac{\that }{\shat },\nonumber \\
^{4}\TGq  & = & -2\frac{T_{-}(\uhat ,\shat )-2\shat (Q^{2}-\shat )}{M_{T}q_{T}}\frac{\that }{\shat }.\label{TqGPP}
\end{eqnarray}

All other non-zero structure functions can be obtained from Eqs.~(\ref{-1TqqUU}-\ref{TqGPP})
by applying the following transformation rules:

\vspace{0.3cm}
\begin{center}\begin{tabular}{ccc}
$\rho =-1,0,1,2$&
$\qquad \qquad $&
$\rho =3,4$\\
\hline
${}^{\rho }T_{\bar{q}q}=\pm ^{\rho }T_{q\bar{q}}$&
&
${}^{\rho }T_{\bar{q}q}=\mp ^{\rho }T_{q\bar{q}}$\\
${}^{\rho }T_{\bar{q}G}=\pm ^{\rho }T_{qG}$&
&
${}^{\rho }T_{\bar{q}G}=\mp ^{\rho }T_{qG}$\\
${}^{\rho }T_{G\bar{q}}=\pm ^{\rho }T_{Gq}$&
&
${}^{\rho }T_{G\bar{q}}=\mp ^{\rho }T_{Gq}$\\
\end{tabular}\end{center}
\vspace{0.3cm}

The upper sign in these rules should be used to obtain the structure
functions in the unpolarized and double-spin cross sections. The lower
sign should be used to obtain the structure functions in the single-spin
cross section. An additional rule should be applied to the unpolarized
and double-spin cross sections in order to relate the structure functions
in the $qG$ and $Gq$ subprocesses:

\begin{tabular}{ccc}
$\rho =-1,0,2,4$&
$\qquad \qquad $&
$\rho =1,3$\\
\hline
${}^{\rho }T_{qG}={}^{\rho }T_{Gq}(\uhat \leftrightarrow \that )$&
&
${}^{\rho }T_{qG}=-^{\rho }T_{Gq}(\uhat \leftrightarrow \that )$\\
\end{tabular}


\begin{thebibliography}{91}
\expandafter\ifx\csname natexlab\endcsname\relax\def\natexlab#1{#1}\fi
\expandafter\ifx\csname bibnamefont\endcsname\relax
  \def\bibnamefont#1{#1}\fi
\expandafter\ifx\csname bibfnamefont\endcsname\relax
  \def\bibfnamefont#1{#1}\fi
\expandafter\ifx\csname citenamefont\endcsname\relax
  \def\citenamefont#1{#1}\fi
\expandafter\ifx\csname url\endcsname\relax
  \def\url#1{\texttt{#1}}\fi
\expandafter\ifx\csname urlprefix\endcsname\relax\def\urlprefix{URL }\fi
\providecommand{\bibinfo}[2]{#2}
\providecommand{\eprint}[2][]{\url{#2}}

\bibitem[{\citenamefont{Boer et~al.}(2001)}]{RHIC2001}
\bibinfo{author}{\bibfnamefont{D.}~\bibnamefont{Boer}} \bibnamefont{et~al.},
  \emph{\bibinfo{title}{\protect{Presentation at the APS Division of Nuclear
  Physics Town Meeting on Nuclear Matter and Hadrons at High Energies}}},
  \bibinfo{howpublished}{Brookhaven National Laboratory}
  (\bibinfo{year}{2001}),
  \bibinfo{note}{http://www.bnl.gov/rhic/townmeeting/bnl.pdf}.

\bibitem[{\citenamefont{Ashman et~al.}(1988)}]{EMC1}
\bibinfo{author}{\bibfnamefont{J.}~\bibnamefont{Ashman}} \bibnamefont{et~al.}
  (\bibinfo{collaboration}{European Muon Collaboration}),
  \bibinfo{journal}{Phys. Lett.} \textbf{\bibinfo{volume}{B206}},
  \bibinfo{pages}{364} (\bibinfo{year}{1988}).

\bibitem[{\citenamefont{Ashman et~al.}(1989)}]{EMC2}
\bibinfo{author}{\bibfnamefont{J.}~\bibnamefont{Ashman}} \bibnamefont{et~al.}
  (\bibinfo{collaboration}{European Muon Collaboration}),
  \bibinfo{journal}{Nucl. Phys.} \textbf{\bibinfo{volume}{B328}},
  \bibinfo{pages}{1} (\bibinfo{year}{1989}).

\bibitem[{\citenamefont{Adeva et~al.}(1998)}]{SMC}
\bibinfo{author}{\bibfnamefont{B.}~\bibnamefont{Adeva}} \bibnamefont{et~al.}
  (\bibinfo{collaboration}{Spin Muon Collaboration}), \bibinfo{journal}{Phys.
  Rev.} \textbf{\bibinfo{volume}{D58}}, \bibinfo{pages}{112001}
  (\bibinfo{year}{1998}).

\bibitem[{\citenamefont{Anthony et~al.}(1996)}]{E142}
\bibinfo{author}{\bibfnamefont{P.~L.} \bibnamefont{Anthony}}
  \bibnamefont{et~al.} (\bibinfo{collaboration}{E142 Collaboration}),
  \bibinfo{journal}{Phys. Rev.} \textbf{\bibinfo{volume}{D54}},
  \bibinfo{pages}{6620} (\bibinfo{year}{1996}).

\bibitem[{\citenamefont{Abe et~al.}(1998)}]{E143}
\bibinfo{author}{\bibfnamefont{K.}~\bibnamefont{Abe}} \bibnamefont{et~al.}
  (\bibinfo{collaboration}{E143 Collaboration}), \bibinfo{journal}{Phys. Rev.}
  \textbf{\bibinfo{volume}{D58}}, \bibinfo{pages}{112003}
  (\bibinfo{year}{1998}).

\bibitem[{\citenamefont{Abe et~al.}(1997)}]{E154}
\bibinfo{author}{\bibfnamefont{K.}~\bibnamefont{Abe}} \bibnamefont{et~al.}
  (\bibinfo{collaboration}{E154 Collaboration}), \bibinfo{journal}{Phys. Rev.
  Lett.} \textbf{\bibinfo{volume}{79}}, \bibinfo{pages}{26}
  (\bibinfo{year}{1997}).

\bibitem[{\citenamefont{Anthony et~al.}(1999)}]{E155}
\bibinfo{author}{\bibfnamefont{P.~L.} \bibnamefont{Anthony}}
  \bibnamefont{et~al.} (\bibinfo{collaboration}{E155 Collaboration}),
  \bibinfo{journal}{Phys. Lett.} \textbf{\bibinfo{volume}{B463}},
  \bibinfo{pages}{339} (\bibinfo{year}{1999}).

\bibitem[{\citenamefont{Ackerstaff et~al.}(1997)}]{HERMES:polDIS96}
\bibinfo{author}{\bibfnamefont{K.}~\bibnamefont{Ackerstaff}}
  \bibnamefont{et~al.} (\bibinfo{collaboration}{HERMES Collaboration}),
  \bibinfo{journal}{Phys. Lett.} \textbf{\bibinfo{volume}{B404}},
  \bibinfo{pages}{383} (\bibinfo{year}{1997}).

\bibitem[{\citenamefont{Airapetian et~al.}(1998)}]{HERMES:polDIS98}
\bibinfo{author}{\bibfnamefont{A.}~\bibnamefont{Airapetian}}
  \bibnamefont{et~al.} (\bibinfo{collaboration}{HERMES Collaboration}),
  \bibinfo{journal}{Phys. Lett.} \textbf{\bibinfo{volume}{B442}},
  \bibinfo{pages}{484} (\bibinfo{year}{1998}).

\bibitem[{\citenamefont{Bunce et~al.}(2000)\citenamefont{Bunce, Saito, Soffer,
  and Vogelsang}}]{RHICOverview}
\bibinfo{author}{\bibfnamefont{G.}~\bibnamefont{Bunce}},
  \bibinfo{author}{\bibfnamefont{N.}~\bibnamefont{Saito}},
  \bibinfo{author}{\bibfnamefont{J.}~\bibnamefont{Soffer}}, \bibnamefont{and}
  \bibinfo{author}{\bibfnamefont{W.}~\bibnamefont{Vogelsang}},
  \bibinfo{journal}{Ann. Rev. Nucl. Part. Sci.} \textbf{\bibinfo{volume}{50}},
  \bibinfo{pages}{525} (\bibinfo{year}{2000}).

\bibitem[{\citenamefont{Close and Sivers}(1977)}]{Close:1977qx}
\bibinfo{author}{\bibfnamefont{F.~E.} \bibnamefont{Close}} \bibnamefont{and}
  \bibinfo{author}{\bibfnamefont{D.~W.} \bibnamefont{Sivers}},
  \bibinfo{journal}{Phys. Rev. Lett.} \textbf{\bibinfo{volume}{39}},
  \bibinfo{pages}{1116} (\bibinfo{year}{1977}).

\bibitem[{\citenamefont{Bourrely et~al.}(1980)\citenamefont{Bourrely, Soffer,
  and Leader}}]{Bourrely:1980mr}
\bibinfo{author}{\bibfnamefont{C.}~\bibnamefont{Bourrely}},
  \bibinfo{author}{\bibfnamefont{J.}~\bibnamefont{Soffer}}, \bibnamefont{and}
  \bibinfo{author}{\bibfnamefont{E.}~\bibnamefont{Leader}},
  \bibinfo{journal}{Phys. Rept.} \textbf{\bibinfo{volume}{59}},
  \bibinfo{pages}{95} (\bibinfo{year}{1980}).

\bibitem[{\citenamefont{Craigie et~al.}(1983)\citenamefont{Craigie, Hidaka,
  Jacob, and Renard}}]{Craigie:1983tk}
\bibinfo{author}{\bibfnamefont{N.~S.} \bibnamefont{Craigie}},
  \bibinfo{author}{\bibfnamefont{K.}~\bibnamefont{Hidaka}},
  \bibinfo{author}{\bibfnamefont{M.}~\bibnamefont{Jacob}}, \bibnamefont{and}
  \bibinfo{author}{\bibfnamefont{F.~M.} \bibnamefont{Renard}},
  \bibinfo{journal}{Phys. Rept.} \textbf{\bibinfo{volume}{99}},
  \bibinfo{pages}{69} (\bibinfo{year}{1983}).

\bibitem[{\citenamefont{Ratcliffe}(1983)}]{Ratcliffe:1983yj}
\bibinfo{author}{\bibfnamefont{P.}~\bibnamefont{Ratcliffe}},
  \bibinfo{journal}{Nucl. Phys.} \textbf{\bibinfo{volume}{B223}},
  \bibinfo{pages}{45} (\bibinfo{year}{1983}).

\bibitem[{\citenamefont{Richter-Was and Szwed}(1985)}]{Richter-Was:1985jp}
\bibinfo{author}{\bibfnamefont{E.}~\bibnamefont{Richter-Was}} \bibnamefont{and}
  \bibinfo{author}{\bibfnamefont{J.}~\bibnamefont{Szwed}},
  \bibinfo{journal}{Phys. Rev.} \textbf{\bibinfo{volume}{D31}},
  \bibinfo{pages}{633} (\bibinfo{year}{1985}).

\bibitem[{\citenamefont{Cheng and Lai}(1990)}]{Cheng:1990hf}
\bibinfo{author}{\bibfnamefont{H.-Y.} \bibnamefont{Cheng}} \bibnamefont{and}
  \bibinfo{author}{\bibfnamefont{S.-N.} \bibnamefont{Lai}},
  \bibinfo{journal}{Phys. Rev.} \textbf{\bibinfo{volume}{D41}},
  \bibinfo{pages}{91} (\bibinfo{year}{1990}).

\bibitem[{\citenamefont{Bourrely et~al.}(1991)\citenamefont{Bourrely, Guillet,
  and Soffer}}]{Bourrely:1991pz}
\bibinfo{author}{\bibfnamefont{C.}~\bibnamefont{Bourrely}},
  \bibinfo{author}{\bibfnamefont{J.~P.} \bibnamefont{Guillet}},
  \bibnamefont{and} \bibinfo{author}{\bibfnamefont{J.}~\bibnamefont{Soffer}},
  \bibinfo{journal}{Nucl. Phys.} \textbf{\bibinfo{volume}{B361}},
  \bibinfo{pages}{72} (\bibinfo{year}{1991}).

\bibitem[{\citenamefont{Weber}(1992)}]{Weber1}
\bibinfo{author}{\bibfnamefont{A.}~\bibnamefont{Weber}},
  \bibinfo{journal}{Nucl. Phys.} \textbf{\bibinfo{volume}{B382}},
  \bibinfo{pages}{63} (\bibinfo{year}{1992}).

\bibitem[{\citenamefont{Mathews and Ravindran}(1992)}]{MathewsRavindran}
\bibinfo{author}{\bibfnamefont{P.}~\bibnamefont{Mathews}} \bibnamefont{and}
  \bibinfo{author}{\bibfnamefont{V.}~\bibnamefont{Ravindran}},
  \bibinfo{journal}{Mod. Phys. Lett.} \textbf{\bibinfo{volume}{A7}},
  \bibinfo{pages}{2695} (\bibinfo{year}{1992}).

\bibitem[{\citenamefont{Chiappetta et~al.}(1993)\citenamefont{Chiappetta,
  Colangelo, Guillet, and Nardulli}}]{Chiappetta:1993cp}
\bibinfo{author}{\bibfnamefont{P.}~\bibnamefont{Chiappetta}},
  \bibinfo{author}{\bibfnamefont{P.}~\bibnamefont{Colangelo}},
  \bibinfo{author}{\bibfnamefont{J.~P.} \bibnamefont{Guillet}},
  \bibnamefont{and} \bibinfo{author}{\bibfnamefont{G.}~\bibnamefont{Nardulli}},
  \bibinfo{journal}{Z. Phys.} \textbf{\bibinfo{volume}{C59}},
  \bibinfo{pages}{629} (\bibinfo{year}{1993}).

\bibitem[{\citenamefont{Kamal}(1996)}]{Kamal1}
\bibinfo{author}{\bibfnamefont{B.}~\bibnamefont{Kamal}},
  \bibinfo{journal}{Phys. Rev.} \textbf{\bibinfo{volume}{D53}},
  \bibinfo{pages}{1142} (\bibinfo{year}{1996}).

\bibitem[{\citenamefont{Gehrmann}(1997)}]{Gehrmann1}
\bibinfo{author}{\bibfnamefont{T.}~\bibnamefont{Gehrmann}},
  \bibinfo{journal}{Nucl. Phys.} \textbf{\bibinfo{volume}{B498}},
  \bibinfo{pages}{245} (\bibinfo{year}{1997}).

\bibitem{Kumano:1999bt}
S.~Kumano and M.~Miyama,
Phys.\ Lett.\ B {\bf 479}, 149 (2000).

\bibitem{Dressler:1999zv}
B.~Dressler, K.~Goeke, M.~V.~Polyakov, P.~Schweitzer, M.~Strikman and C.~Weiss,
Eur.\ Phys.\ J.\ C {\bf 18}, 719 (2001).

\bibitem{Gluck:2000ek}
M.~Gluck, A.~Hartl and E.~Reya,
Eur.\ Phys.\ J.\ C {\bf 19}, 77 (2001).

\bibitem[{\citenamefont{Ravindran et~al.}(2002)\citenamefont{Ravindran, Smith,
  and van Neerven}}]{Ravindran:2002na}
\bibinfo{author}{\bibfnamefont{V.}~\bibnamefont{Ravindran}},
  \bibinfo{author}{\bibfnamefont{J.}~\bibnamefont{Smith}}, \bibnamefont{and}
  \bibinfo{author}{\bibfnamefont{W.~L.} \bibnamefont{van Neerven}},
  \bibinfo{journal}{Nucl. Phys.} \textbf{\bibinfo{volume}{B647}},
  \bibinfo{pages}{275} (\bibinfo{year}{2002}).

\bibitem{Kodaira:2003tq}
J.~Kodaira and H.~Yokoya,
Phys.\ Rev.\ D {\bf 67}, 074008 (2003).

\bibitem[{\citenamefont{Chiappetta and Soffer}(1985)}]{Chiappetta:1985dy}
\bibinfo{author}{\bibfnamefont{P.}~\bibnamefont{Chiappetta}} \bibnamefont{and}
  \bibinfo{author}{\bibfnamefont{J.}~\bibnamefont{Soffer}},
  \bibinfo{journal}{Phys. Lett.} \textbf{\bibinfo{volume}{B152}},
  \bibinfo{pages}{126} (\bibinfo{year}{1985}).

\bibitem[{\citenamefont{Weber}(1993)}]{Weber2}
\bibinfo{author}{\bibfnamefont{A.}~\bibnamefont{Weber}},
  \bibinfo{journal}{Nucl. Phys.} \textbf{\bibinfo{volume}{B403}},
  \bibinfo{pages}{545} (\bibinfo{year}{1993}).

\bibitem[{\citenamefont{Bourrely and Soffer}(1994)}]{Bourrely:1994sc}
\bibinfo{author}{\bibfnamefont{C.}~\bibnamefont{Bourrely}} \bibnamefont{and}
  \bibinfo{author}{\bibfnamefont{J.}~\bibnamefont{Soffer}},
  \bibinfo{journal}{Nucl. Phys.} \textbf{\bibinfo{volume}{B423}},
  \bibinfo{pages}{329} (\bibinfo{year}{1994}).

\bibitem[{\citenamefont{Bourrely and Soffer}(1993)}]{Bourrely:1993dd}
\bibinfo{author}{\bibfnamefont{C.}~\bibnamefont{Bourrely}} \bibnamefont{and}
  \bibinfo{author}{\bibfnamefont{J.}~\bibnamefont{Soffer}},
  \bibinfo{journal}{Phys. Lett.} \textbf{\bibinfo{volume}{B314}},
  \bibinfo{pages}{132} (\bibinfo{year}{1993}).

\bibitem[{\citenamefont{Nadolsky}(1995)}]{Nadolsky:1995nf}
\bibinfo{author}{\bibfnamefont{P.~M.} \bibnamefont{Nadolsky}}
  (\bibinfo{year}{1995}), \eprint{hep-ph/9503419}.

\bibitem[{\citenamefont{Bourrely and Soffer}(1995)}]{Bourrely:1995fw}
\bibinfo{author}{\bibfnamefont{C.}~\bibnamefont{Bourrely}} \bibnamefont{and}
  \bibinfo{author}{\bibfnamefont{J.}~\bibnamefont{Soffer}},
  \bibinfo{journal}{Nucl. Phys.} \textbf{\bibinfo{volume}{B445}},
  \bibinfo{pages}{341} (\bibinfo{year}{1995}).

\bibitem[{\citenamefont{Kamal}(1998)}]{Kamal2}
\bibinfo{author}{\bibfnamefont{B.}~\bibnamefont{Kamal}},
  \bibinfo{journal}{Phys. Rev.} \textbf{\bibinfo{volume}{D57}},
  \bibinfo{pages}{6663} (\bibinfo{year}{1998}).

\bibitem[{\citenamefont{Gehrmann}(1998)}]{Gehrmann2}
\bibinfo{author}{\bibfnamefont{T.}~\bibnamefont{Gehrmann}},
  \bibinfo{journal}{Nucl. Phys.} \textbf{\bibinfo{volume}{B534}},
  \bibinfo{pages}{21} (\bibinfo{year}{1998}).

\bibitem[{\citenamefont{Cabibbo}(1963)}]{Cabibbo:1963yz}
\bibinfo{author}{\bibfnamefont{N.}~\bibnamefont{Cabibbo}},
  \bibinfo{journal}{Phys. Rev. Lett.} \textbf{\bibinfo{volume}{10}},
  \bibinfo{pages}{531} (\bibinfo{year}{1963}).

\bibitem[{\citenamefont{Kobayashi and Maskawa}(1973)}]{Kobayashi:1973fv}
\bibinfo{author}{\bibfnamefont{M.}~\bibnamefont{Kobayashi}} \bibnamefont{and}
  \bibinfo{author}{\bibfnamefont{T.}~\bibnamefont{Maskawa}},
  \bibinfo{journal}{Prog. Theor. Phys.} \textbf{\bibinfo{volume}{49}},
  \bibinfo{pages}{652} (\bibinfo{year}{1973}).

\bibitem[{\citenamefont{Ackerstaff et~al.}(1999)}]{HERMES:polSIDIS99}
\bibinfo{author}{\bibfnamefont{K.}~\bibnamefont{Ackerstaff}}
  \bibnamefont{et~al.} (\bibinfo{collaboration}{HERMES Collaboration}),
  \bibinfo{journal}{Phys. Lett.} \textbf{\bibinfo{volume}{B464}},
  \bibinfo{pages}{123} (\bibinfo{year}{1999}).

\bibitem[{\citenamefont{Bland}(2000)}]{Bland:1999gb}
\bibinfo{author}{\bibfnamefont{L.~C.} \bibnamefont{Bland}}
  (\bibinfo{collaboration}{STAR Collaboration}) (\bibinfo{year}{2000}),
  \eprint{hep-ex/0002061}.

\bibitem[{\citenamefont{Nadolsky and Yuan}(2003)}]{PolWPhenomenology}
\bibinfo{author}{\bibfnamefont{P.~M.} \bibnamefont{Nadolsky}} \bibnamefont{and}
  \bibinfo{author}{\bibfnamefont{C.-P.} \bibnamefont{Yuan}}
  (\bibinfo{year}{2003}), 
  \eprint{hep-ph/0304002}.

\bibitem[{\citenamefont{Dokshitzer et~al.}(1978)\citenamefont{Dokshitzer,
  Diakonov, and Troian}}]{DDT78}
\bibinfo{author}{\bibfnamefont{Y.~L.} \bibnamefont{Dokshitzer}},
  \bibinfo{author}{\bibfnamefont{D.}~\bibnamefont{Diakonov}}, \bibnamefont{and}
  \bibinfo{author}{\bibfnamefont{S.~I.} \bibnamefont{Troian}},
  \bibinfo{journal}{Phys. Lett.} \textbf{\bibinfo{volume}{79B}},
  \bibinfo{pages}{269} (\bibinfo{year}{1978}).

\bibitem[{\citenamefont{Parisi and Petronzio}(1979)}]{ParisiPetronzio79}
\bibinfo{author}{\bibfnamefont{G.}~\bibnamefont{Parisi}} \bibnamefont{and}
  \bibinfo{author}{\bibfnamefont{R.}~\bibnamefont{Petronzio}},
  \bibinfo{journal}{Nucl. Phys.} \textbf{\bibinfo{volume}{B154}},
  \bibinfo{pages}{427} (\bibinfo{year}{1979}).

\bibitem[{\citenamefont{Altarelli et~al.}(1979)\citenamefont{Altarelli, Ellis,
  and Martinelli}}]{AltarelliEllisMartinelli79}
\bibinfo{author}{\bibfnamefont{G.}~\bibnamefont{Altarelli}},
  \bibinfo{author}{\bibfnamefont{R.~K.} \bibnamefont{Ellis}}, \bibnamefont{and}
  \bibinfo{author}{\bibfnamefont{G.}~\bibnamefont{Martinelli}},
  \bibinfo{journal}{Nucl. Phys.} \textbf{\bibinfo{volume}{B157}},
  \bibinfo{pages}{461} (\bibinfo{year}{1979}).

\bibitem[{\citenamefont{Collins and Soper}(1981)}]{CollinsSoper81}
\bibinfo{author}{\bibfnamefont{J.~C.} \bibnamefont{Collins}} \bibnamefont{and}
  \bibinfo{author}{\bibfnamefont{D.~E.} \bibnamefont{Soper}},
  \bibinfo{journal}{Nucl. Phys.} \textbf{\bibinfo{volume}{B193}},
  \bibinfo{pages}{381} (\bibinfo{year}{1981}).

\bibitem[{\citenamefont{Collins and Soper}(1982)}]{CollinsSoper82}
\bibinfo{author}{\bibfnamefont{J.~C.} \bibnamefont{Collins}} \bibnamefont{and}
  \bibinfo{author}{\bibfnamefont{D.~E.} \bibnamefont{Soper}},
  \bibinfo{journal}{Nucl. Phys.} \textbf{\bibinfo{volume}{B197}},
  \bibinfo{pages}{446} (\bibinfo{year}{1982}).

\bibitem[{\citenamefont{Collins et~al.}(1985)\citenamefont{Collins, Soper, and
  Sterman}}]{CSS}
\bibinfo{author}{\bibfnamefont{J.~C.} \bibnamefont{Collins}},
  \bibinfo{author}{\bibfnamefont{D.~E.} \bibnamefont{Soper}}, \bibnamefont{and}
  \bibinfo{author}{\bibfnamefont{G.}~\bibnamefont{Sterman}},
  \bibinfo{journal}{Nucl. Phys.} \textbf{\bibinfo{volume}{B250}},
  \bibinfo{pages}{199} (\bibinfo{year}{1985}).

\bibitem[{\citenamefont{Davies et~al.}(1985)\citenamefont{Davies, Webber, and
  Stirling}}]{DWS}
\bibinfo{author}{\bibfnamefont{C.~T.~H.} \bibnamefont{Davies}},
  \bibinfo{author}{\bibfnamefont{B.~R.} \bibnamefont{Webber}},
  \bibnamefont{and} \bibinfo{author}{\bibfnamefont{W.~J.}
  \bibnamefont{Stirling}}, \bibinfo{journal}{Nucl. Phys.}
  \textbf{\bibinfo{volume}{B256}}, \bibinfo{pages}{413} (\bibinfo{year}{1985}).

\bibitem[{\citenamefont{Arnold and Kauffman}(1991)}]{Arnold:1991yk}
\bibinfo{author}{\bibfnamefont{P.~B.} \bibnamefont{Arnold}} \bibnamefont{and}
  \bibinfo{author}{\bibfnamefont{R.~P.} \bibnamefont{Kauffman}},
  \bibinfo{journal}{Nucl. Phys.} \textbf{\bibinfo{volume}{B349}},
  \bibinfo{pages}{381} (\bibinfo{year}{1991}).

\bibitem[{\citenamefont{Balazs and Yuan}(1997)}]{CPCsaba}
\bibinfo{author}{\bibfnamefont{C.}~\bibnamefont{Balazs}} \bibnamefont{and}
  \bibinfo{author}{\bibfnamefont{C.-P.} \bibnamefont{Yuan}},
  \bibinfo{journal}{Phys. Rev.} \textbf{\bibinfo{volume}{D56}},
  \bibinfo{pages}{5558} (\bibinfo{year}{1997}).

\bibitem[{\citenamefont{Ladinsky and Yuan}(1994)}]{Ladinsky:1994zn}
\bibinfo{author}{\bibfnamefont{G.~A.} \bibnamefont{Ladinsky}} \bibnamefont{and}
  \bibinfo{author}{\bibfnamefont{C.-P.} \bibnamefont{Yuan}},
  \bibinfo{journal}{Phys. Rev.} \textbf{\bibinfo{volume}{D50}},
  \bibinfo{pages}{4239} (\bibinfo{year}{1994}).

\bibitem[{\citenamefont{Ellis et~al.}(1997)\citenamefont{Ellis, Ross, and
  Veseli}}]{Ellis:1997sc}
\bibinfo{author}{\bibfnamefont{R.~K.} \bibnamefont{Ellis}},
  \bibinfo{author}{\bibfnamefont{D.~A.} \bibnamefont{Ross}}, \bibnamefont{and}
  \bibinfo{author}{\bibfnamefont{S.}~\bibnamefont{Veseli}},
  \bibinfo{journal}{Nucl. Phys.} \textbf{\bibinfo{volume}{B503}},
  \bibinfo{pages}{309} (\bibinfo{year}{1997}).

\bibitem[{\citenamefont{Ellis and Veseli}(1998)}]{Ellis:1998ii}
\bibinfo{author}{\bibfnamefont{R.~K.} \bibnamefont{Ellis}} \bibnamefont{and}
  \bibinfo{author}{\bibfnamefont{S.}~\bibnamefont{Veseli}},
  \bibinfo{journal}{Nucl. Phys.} \textbf{\bibinfo{volume}{B511}},
  \bibinfo{pages}{649} (\bibinfo{year}{1998}).

\bibitem[{\citenamefont{Landry et~al.}(2001)\citenamefont{Landry, Brock,
  Ladinsky, and Yuan}}]{Landry:1999an}
\bibinfo{author}{\bibfnamefont{F.}~\bibnamefont{Landry}},
  \bibinfo{author}{\bibfnamefont{R.}~\bibnamefont{Brock}},
  \bibinfo{author}{\bibfnamefont{G.}~\bibnamefont{Ladinsky}}, \bibnamefont{and}
  \bibinfo{author}{\bibfnamefont{C.-P.} \bibnamefont{Yuan}},
  \bibinfo{journal}{Phys. Rev.} \textbf{\bibinfo{volume}{D63}},
  \bibinfo{pages}{013004} (\bibinfo{year}{2001}).

\bibitem[{\citenamefont{Qiu and Zhang}(2001{\natexlab{a}})}]{Qiu:2000ga}
\bibinfo{author}{\bibfnamefont{J.}~\bibnamefont{Qiu}} \bibnamefont{and}
  \bibinfo{author}{\bibfnamefont{X.}~\bibnamefont{Zhang}},
  \bibinfo{journal}{Phys. Rev. Lett.} \textbf{\bibinfo{volume}{86}},
  \bibinfo{pages}{2724} (\bibinfo{year}{2001}{\natexlab{a}}).

\bibitem[{\citenamefont{Qiu and Zhang}(2001{\natexlab{b}})}]{Qiu:2000hf}
\bibinfo{author}{\bibfnamefont{J.}~\bibnamefont{Qiu}} \bibnamefont{and}
  \bibinfo{author}{\bibfnamefont{X.}~\bibnamefont{Zhang}},
  \bibinfo{journal}{Phys. Rev.} \textbf{\bibinfo{volume}{D63}},
  \bibinfo{pages}{114011} (\bibinfo{year}{2001}{\natexlab{b}}).

\bibitem[{\citenamefont{Kulesza and Stirling}(2001)}]{Kulesza:2001jc}
\bibinfo{author}{\bibfnamefont{A.}~\bibnamefont{Kulesza}} \bibnamefont{and}
  \bibinfo{author}{\bibfnamefont{W.~J.} \bibnamefont{Stirling}},
  \bibinfo{journal}{Eur. Phys. J.} \textbf{\bibinfo{volume}{C20}},
  \bibinfo{pages}{349} (\bibinfo{year}{2001}).

\bibitem[{\citenamefont{Landry et~al.}(2002)\citenamefont{Landry, Brock,
  Nadolsky, and Yuan}}]{Landry:2002ix}
\bibinfo{author}{\bibfnamefont{F.}~\bibnamefont{Landry}},
  \bibinfo{author}{\bibfnamefont{R.}~\bibnamefont{Brock}},
  \bibinfo{author}{\bibfnamefont{P.~M.} \bibnamefont{Nadolsky}},
  \bibnamefont{and} \bibinfo{author}{\bibfnamefont{C.-P.} \bibnamefont{Yuan}}
  (\bibinfo{year}{2002}), \eprint{hep-ph/0212159}.

\bibitem[{\citenamefont{Siegel}(1979)}]{Siegel:1979wq}
\bibinfo{author}{\bibfnamefont{W.}~\bibnamefont{Siegel}},
  \bibinfo{journal}{Phys. Lett.} \textbf{\bibinfo{volume}{84B}},
  \bibinfo{pages}{193} (\bibinfo{year}{1979}).

\bibitem[{\citenamefont{Schuler et~al.}(1987)\citenamefont{Schuler, Sakakibara,
  and Korner}}]{Schuler:1987ej}
\bibinfo{author}{\bibfnamefont{G.~A.} \bibnamefont{Schuler}},
  \bibinfo{author}{\bibfnamefont{S.}~\bibnamefont{Sakakibara}},
  \bibnamefont{and} \bibinfo{author}{\bibfnamefont{J.~G.}
  \bibnamefont{Korner}}, \bibinfo{journal}{Phys. Lett.}
  \textbf{\bibinfo{volume}{B194}}, \bibinfo{pages}{125} (\bibinfo{year}{1987}).

\bibitem[{\citenamefont{Korner and Tung}(1994)}]{Korner:1994pv}
\bibinfo{author}{\bibfnamefont{J.~G.} \bibnamefont{Korner}} \bibnamefont{and}
  \bibinfo{author}{\bibfnamefont{M.~M.} \bibnamefont{Tung}},
  \bibinfo{journal}{Z. Phys.} \textbf{\bibinfo{volume}{C64}},
  \bibinfo{pages}{255} (\bibinfo{year}{1994}).

\bibitem[{\citenamefont{'t~Hooft and Veltman}(1972)}]{'tHooft:1972fi}
\bibinfo{author}{\bibfnamefont{G.}~\bibnamefont{'t~Hooft}} \bibnamefont{and}
  \bibinfo{author}{\bibfnamefont{M.~J.~G.} \bibnamefont{Veltman}},
  \bibinfo{journal}{Nucl. Phys.} \textbf{\bibinfo{volume}{B44}},
  \bibinfo{pages}{189} (\bibinfo{year}{1972}).

\bibitem[{\citenamefont{Breitenlohner and Maison}(1977)}]{Breitenlohner:1977hr}
\bibinfo{author}{\bibfnamefont{P.}~\bibnamefont{Breitenlohner}}
  \bibnamefont{and} \bibinfo{author}{\bibfnamefont{D.}~\bibnamefont{Maison}},
  \bibinfo{journal}{Commun. Math. Phys.} \textbf{\bibinfo{volume}{52}},
  \bibinfo{pages}{11} (\bibinfo{year}{1977}).

\bibitem[{\citenamefont{Collins and Soper}(1977)}]{CSframe}
\bibinfo{author}{\bibfnamefont{J.~C.} \bibnamefont{Collins}} \bibnamefont{and}
  \bibinfo{author}{\bibfnamefont{D.~E.} \bibnamefont{Soper}},
  \bibinfo{journal}{Phys. Rev.} \textbf{\bibinfo{volume}{D16}},
  \bibinfo{pages}{2219} (\bibinfo{year}{1977}).

\bibitem[{\citenamefont{Pire and Ralston}(1983)}]{Pire:1983tv}
\bibinfo{author}{\bibfnamefont{B.}~\bibnamefont{Pire}} \bibnamefont{and}
  \bibinfo{author}{\bibfnamefont{J.~P.} \bibnamefont{Ralston}},
  \bibinfo{journal}{Phys. Rev.} \textbf{\bibinfo{volume}{D28}},
  \bibinfo{pages}{260} (\bibinfo{year}{1983}).

\bibitem[{\citenamefont{Carlitz and Willey}(1992)}]{Carlitz:1992fv}
\bibinfo{author}{\bibfnamefont{R.~D.} \bibnamefont{Carlitz}} \bibnamefont{and}
  \bibinfo{author}{\bibfnamefont{R.~S.} \bibnamefont{Willey}},
  \bibinfo{journal}{Phys. Rev.} \textbf{\bibinfo{volume}{D45}},
  \bibinfo{pages}{2323} (\bibinfo{year}{1992}).

\bibitem[{\citenamefont{Nadolsky}(1994)}]{Nadolsky:1994uq}
\bibinfo{author}{\bibfnamefont{P.~M.} \bibnamefont{Nadolsky}},
  \bibinfo{journal}{Z. Phys.} \textbf{\bibinfo{volume}{C62}},
  \bibinfo{pages}{109} (\bibinfo{year}{1994}).

\bibitem[{\citenamefont{Bollini and
  Giambiagi}(1972{\natexlab{a}})}]{Bollini:1972ui}
\bibinfo{author}{\bibfnamefont{C.~G.} \bibnamefont{Bollini}} \bibnamefont{and}
  \bibinfo{author}{\bibfnamefont{J.~J.} \bibnamefont{Giambiagi}},
  \bibinfo{journal}{Nuovo Cim.} \textbf{\bibinfo{volume}{B12}},
  \bibinfo{pages}{20} (\bibinfo{year}{1972}{\natexlab{a}}).

\bibitem[{\citenamefont{Bollini and
  Giambiagi}(1972{\natexlab{b}})}]{Bollini:1972bi}
\bibinfo{author}{\bibfnamefont{C.~G.} \bibnamefont{Bollini}} \bibnamefont{and}
  \bibinfo{author}{\bibfnamefont{J.~J.} \bibnamefont{Giambiagi}},
  \bibinfo{journal}{Phys. Lett.} \textbf{\bibinfo{volume}{B40}},
  \bibinfo{pages}{566} (\bibinfo{year}{1972}{\natexlab{b}}).

\bibitem[{Col()}]{CollinsRenormalizationBook}
\bibinfo{howpublished}{{\protect See, e.g., J. C. Collins, {\it
  Renormalization}, ch. 13 (Cambridge University Press, 1984)}}.

\bibitem[{\citenamefont{Capper et~al.}(1980)\citenamefont{Capper, Jones, and
  van Nieuwenhuizen}}]{Capper:1980ns}
\bibinfo{author}{\bibfnamefont{D.~M.} \bibnamefont{Capper}},
  \bibinfo{author}{\bibfnamefont{D.~R.~T.} \bibnamefont{Jones}},
  \bibnamefont{and} \bibinfo{author}{\bibfnamefont{P.}~\bibnamefont{van
  Nieuwenhuizen}}, \bibinfo{journal}{Nucl. Phys.}
  \textbf{\bibinfo{volume}{B167}}, \bibinfo{pages}{479} (\bibinfo{year}{1980}).

\bibitem[{\citenamefont{Avdeev et~al.}(1980)\citenamefont{Avdeev, Tarasov, and
  Vladimirov}}]{Avdeev:1980bh}
\bibinfo{author}{\bibfnamefont{L.~V.} \bibnamefont{Avdeev}},
  \bibinfo{author}{\bibfnamefont{O.~V.} \bibnamefont{Tarasov}},
  \bibnamefont{and} \bibinfo{author}{\bibfnamefont{A.~A.}
  \bibnamefont{Vladimirov}}, \bibinfo{journal}{Phys. Lett.}
  \textbf{\bibinfo{volume}{B96}}, \bibinfo{pages}{94} (\bibinfo{year}{1980}).

\bibitem[{\citenamefont{Korner et~al.}(1992)\citenamefont{Korner, Kreimer, and
  Schilcher}}]{Korner:1992sx}
\bibinfo{author}{\bibfnamefont{J.~G.} \bibnamefont{Korner}},
  \bibinfo{author}{\bibfnamefont{D.}~\bibnamefont{Kreimer}}, \bibnamefont{and}
  \bibinfo{author}{\bibfnamefont{K.}~\bibnamefont{Schilcher}},
  \bibinfo{journal}{Z. Phys.} \textbf{\bibinfo{volume}{C54}},
  \bibinfo{pages}{503} (\bibinfo{year}{1992}).

\bibitem[{\citenamefont{Chanowitz et~al.}(1978)\citenamefont{Chanowitz, Furman,
  and Hinchliffe}}]{Chanowitz:1978uj}
\bibinfo{author}{\bibfnamefont{M.~S.} \bibnamefont{Chanowitz}},
  \bibinfo{author}{\bibfnamefont{M.~A.} \bibnamefont{Furman}},
  \bibnamefont{and}
  \bibinfo{author}{\bibfnamefont{I.}~\bibnamefont{Hinchliffe}},
  \bibinfo{journal}{Phys. Lett.} \textbf{\bibinfo{volume}{B78}},
  \bibinfo{pages}{285} (\bibinfo{year}{1978}).

\bibitem[{\citenamefont{Chanowitz et~al.}(1979)\citenamefont{Chanowitz, Furman,
  and Hinchliffe}}]{Chanowitz:1979zu}
\bibinfo{author}{\bibfnamefont{M.~S.} \bibnamefont{Chanowitz}},
  \bibinfo{author}{\bibfnamefont{M.}~\bibnamefont{Furman}}, \bibnamefont{and}
  \bibinfo{author}{\bibfnamefont{I.}~\bibnamefont{Hinchliffe}},
  \bibinfo{journal}{Nucl. Phys.} \textbf{\bibinfo{volume}{B159}},
  \bibinfo{pages}{225} (\bibinfo{year}{1979}).

\bibitem[{\citenamefont{Siegel}(1980)}]{Siegel:1980qs}
\bibinfo{author}{\bibfnamefont{W.}~\bibnamefont{Siegel}},
  \bibinfo{journal}{Phys. Lett.} \textbf{\bibinfo{volume}{B94}},
  \bibinfo{pages}{37} (\bibinfo{year}{1980}).

\bibitem[{\citenamefont{Avdeev and Vladimirov}(1983)}]{Avdeev:1983xy}
\bibinfo{author}{\bibfnamefont{L.~V.} \bibnamefont{Avdeev}} \bibnamefont{and}
  \bibinfo{author}{\bibfnamefont{A.~A.} \bibnamefont{Vladimirov}},
  \bibinfo{journal}{Nucl. Phys.} \textbf{\bibinfo{volume}{B219}},
  \bibinfo{pages}{262} (\bibinfo{year}{1983}).

\bibitem[{\citenamefont{Kunszt et~al.}(1994)\citenamefont{Kunszt, Signer, and
  Trocsanyi}}]{Kunszt:1994sd}
\bibinfo{author}{\bibfnamefont{Z.}~\bibnamefont{Kunszt}},
  \bibinfo{author}{\bibfnamefont{A.}~\bibnamefont{Signer}}, \bibnamefont{and}
  \bibinfo{author}{\bibfnamefont{Z.}~\bibnamefont{Trocsanyi}},
  \bibinfo{journal}{Nucl. Phys.} \textbf{\bibinfo{volume}{B411}},
  \bibinfo{pages}{397} (\bibinfo{year}{1994}), \eprint{hep-ph/9305239}.

\bibitem[{\citenamefont{Jamin and Lautenbacher}(1993)}]{Jamin:1993dp}
\bibinfo{author}{\bibfnamefont{M.}~\bibnamefont{Jamin}} \bibnamefont{and}
  \bibinfo{author}{\bibfnamefont{M.~E.} \bibnamefont{Lautenbacher}},
  \bibinfo{journal}{Comput. Phys. Commun.} \textbf{\bibinfo{volume}{74}},
  \bibinfo{pages}{265} (\bibinfo{year}{1993}).

\bibitem[{\citenamefont{Dokshitzer}(1977)}]{Dokshitzer77}
\bibinfo{author}{\bibfnamefont{Y.~L.} \bibnamefont{Dokshitzer}},
  \bibinfo{journal}{Sov. Phys. JETP} \textbf{\bibinfo{volume}{46}},
  \bibinfo{pages}{641} (\bibinfo{year}{1977}),
  \bibinfo{note}{\protect{Zh.Eksp.Teor.Fiz. 73, 1216 (1977)}}.

\bibitem[{\citenamefont{Gribov and Lipatov}(1972)}]{GribovLipatov72}
\bibinfo{author}{\bibfnamefont{V.~N.} \bibnamefont{Gribov}} \bibnamefont{and}
  \bibinfo{author}{\bibfnamefont{L.~N.} \bibnamefont{Lipatov}},
  \bibinfo{journal}{Sov. J. Nucl. Phys.} \textbf{\bibinfo{volume}{15}},
  \bibinfo{pages}{438} (\bibinfo{year}{1972}), \bibinfo{note}{\protect{Yad.Fiz. 15, 781
  (1972)}}.

\bibitem[{\citenamefont{Altarelli and Parisi}(1977)}]{AP}
\bibinfo{author}{\bibfnamefont{G.}~\bibnamefont{Altarelli}} \bibnamefont{and}
  \bibinfo{author}{\bibfnamefont{G.}~\bibnamefont{Parisi}},
  \bibinfo{journal}{Nucl. Phys.} \textbf{\bibinfo{volume}{B126}},
  \bibinfo{pages}{298} (\bibinfo{year}{1977}).

\bibitem[{\citenamefont{Ahmed and Ross}(1976)}]{AhmedRoss}
\bibinfo{author}{\bibfnamefont{M.~A.} \bibnamefont{Ahmed}} \bibnamefont{and}
  \bibinfo{author}{\bibfnamefont{G.~G.} \bibnamefont{Ross}},
  \bibinfo{journal}{Nucl. Phys.} \textbf{\bibinfo{volume}{B111}},
  \bibinfo{pages}{441} (\bibinfo{year}{1976}).

\bibitem[{\citenamefont{Ellis et~al.}(1981)\citenamefont{Ellis, Ross, and
  Terrano}}]{Ellis:1981wv}
\bibinfo{author}{\bibfnamefont{R.~K.} \bibnamefont{Ellis}},
  \bibinfo{author}{\bibfnamefont{D.~A.} \bibnamefont{Ross}}, \bibnamefont{and}
  \bibinfo{author}{\bibfnamefont{A.~E.} \bibnamefont{Terrano}},
  \bibinfo{journal}{Nucl. Phys.} \textbf{\bibinfo{volume}{B178}},
  \bibinfo{pages}{421} (\bibinfo{year}{1981}).

\bibitem[{\citenamefont{Gordon and Vogelsang}(1993)}]{Gordon:1993qc}
\bibinfo{author}{\bibfnamefont{L.~E.} \bibnamefont{Gordon}} \bibnamefont{and}
  \bibinfo{author}{\bibfnamefont{W.}~\bibnamefont{Vogelsang}},
  \bibinfo{journal}{Phys. Rev.} \textbf{\bibinfo{volume}{D48}},
  \bibinfo{pages}{3136} (\bibinfo{year}{1993}).

\bibitem[{\citenamefont{Mertig and van Neerven}(1996)}]{Mertig:1996ny}
\bibinfo{author}{\bibfnamefont{R.}~\bibnamefont{Mertig}} \bibnamefont{and}
  \bibinfo{author}{\bibfnamefont{W.~L.} \bibnamefont{van Neerven}},
  \bibinfo{journal}{Z. Phys.} \textbf{\bibinfo{volume}{C70}},
  \bibinfo{pages}{637} (\bibinfo{year}{1996}).

\bibitem[{\citenamefont{Vogelsang}(1996{\natexlab{a}})}]{Vogelsang:1996vh}
\bibinfo{author}{\bibfnamefont{W.}~\bibnamefont{Vogelsang}},
  \bibinfo{journal}{Phys. Rev.} \textbf{\bibinfo{volume}{D54}},
  \bibinfo{pages}{2023} (\bibinfo{year}{1996}{\natexlab{a}}).

\bibitem[{\citenamefont{Vogelsang}(1996{\natexlab{b}})}]{VogelsangTwoLoop}
\bibinfo{author}{\bibfnamefont{W.}~\bibnamefont{Vogelsang}},
  \bibinfo{journal}{Nucl. Phys.} \textbf{\bibinfo{volume}{B475}},
  \bibinfo{pages}{47} (\bibinfo{year}{1996}{\natexlab{b}}).

\bibitem[{\citenamefont{de~Florian and Grazzini}(2000)}]{deFlorianPRL2000}
\bibinfo{author}{\bibfnamefont{D.}~\bibnamefont{de~Florian}} \bibnamefont{and}
  \bibinfo{author}{\bibfnamefont{M.}~\bibnamefont{Grazzini}},
  \bibinfo{journal}{Phys. Rev. Lett.} \textbf{\bibinfo{volume}{85}},
  \bibinfo{pages}{4678} (\bibinfo{year}{2000}).

\bibitem[{\citenamefont{Catani et~al.}(2001)\citenamefont{Catani, de~Florian,
  and Grazzini}}]{CataniDeFlorianGrazzini2000}
\bibinfo{author}{\bibfnamefont{S.}~\bibnamefont{Catani}},
  \bibinfo{author}{\bibfnamefont{D.}~\bibnamefont{de~Florian}},
  \bibnamefont{and} \bibinfo{author}{\bibfnamefont{M.}~\bibnamefont{Grazzini}},
  \bibinfo{journal}{Nucl. Phys.} \textbf{\bibinfo{volume}{B596}},
  \bibinfo{pages}{299} (\bibinfo{year}{2001}).

\bibitem[{\citenamefont{de~Florian and Grazzini}(2001)}]{deFlorian:2001zd}
\bibinfo{author}{\bibfnamefont{D.}~\bibnamefont{de~Florian}} \bibnamefont{and}
  \bibinfo{author}{\bibfnamefont{M.}~\bibnamefont{Grazzini}},
  \bibinfo{journal}{Nucl. Phys.} \textbf{\bibinfo{volume}{B616}},
  \bibinfo{pages}{247} (\bibinfo{year}{2001}).

\bibitem[{\citenamefont{Kodaira and Trentadue}(1982)}]{KodairaA2}
\bibinfo{author}{\bibfnamefont{J.}~\bibnamefont{Kodaira}} \bibnamefont{and}
  \bibinfo{author}{\bibfnamefont{L.}~\bibnamefont{Trentadue}},
  \bibinfo{journal}{Phys. Lett.} \textbf{\bibinfo{volume}{B112}},
  \bibinfo{pages}{66} (\bibinfo{year}{1982}).

\bibitem[{\citenamefont{Davies and Stirling}(1984)}]{DaviesStirling84}
\bibinfo{author}{\bibfnamefont{C.~T.~H.} \bibnamefont{Davies}}
  \bibnamefont{and} \bibinfo{author}{\bibfnamefont{W.~J.}
  \bibnamefont{Stirling}}, \bibinfo{journal}{Nucl. Phys.}
  \textbf{\bibinfo{volume}{B244}}, \bibinfo{pages}{337} (\bibinfo{year}{1984}).

\end{thebibliography}
\end{document}